\documentclass[traditabstract]{aa}
\usepackage{ifpdf}
\usepackage{times}
\usepackage{longtable}
\usepackage{natbib}
\usepackage{rotate}
\usepackage{lscape}
\usepackage{amssymb}

\ifpdf
  \usepackage[pdftex]{graphicx,color}
\else
  \usepackage[dvips]{graphicx,color}
\fi
\bibpunct{(}{)}{;}{a}{}{,} 

\def\5{\footnotesize V\normalsize}
\def\4{\footnotesize IV\normalsize}
\def\3{\footnotesize III\normalsize}
\def\2{\footnotesize II\normalsize}
\def\1{\footnotesize I\normalsize}
\def\lam{$\lambda$}
\def\kms{$\mbox{km s}^{-1}$}

\def\pp{$\phantom{-}$}
\def\oo{$\phantom{1}$}

\begin{document}

\title{The VLT-FLAMES Tarantula Survey\\ I: Introduction and observational overview}

\author{C.~J.~Evans\inst{1}, W.~D.~Taylor\inst{2}, 
V.~H\'{e}nault-Brunet\inst{2}, H.~Sana\inst{3}, 
A.~de~Koter\inst{3,4}, S.~Sim\'{o}n-D\'{i}az\inst{5,6}, G.~Carraro\inst{7}, \\
T.~Bagnoli\inst{3}, N.~Bastian\inst{8,9},
J.~M.~Bestenlehner\inst{10}, A.~Z.~Bonanos\inst{11}, E.~Bressert\inst{9,12,13}, I.~Brott\inst{4,14}, 
M.~A.~Campbell\inst{2}, M.~Cantiello\inst{15}, J.~S.~Clark\inst{16}, E.~Costa\inst{17}, 
P.~A.~Crowther\inst{18}, S.~E.~de~Mink\inst{19}\thanks{Hubble Fellow}, E.~Doran\inst{18}, P.~L.~Dufton\inst{20},
P.~R.~Dunstall\inst{20}, K.~Friedrich\inst{15}, M.~Garcia\inst{5,6}, M.~Gieles\inst{21}, G.~Gr\"{a}fener\inst{10}, 
A.~Herrero\inst{5,6}, I.~D.~Howarth\inst{22}, R.~G.~Izzard\inst{15}, N.~Langer\inst{15},
D.~J.~Lennon\inst{23}, J.~Ma\'{i}z~Apell\'{a}niz\inst{24}\thanks{Ram\'on y Cajal Fellow}, N.~Markova\inst{25}, F.~Najarro\inst{26}, J.~Puls\inst{27},
O.~H.~Ramirez\inst{3}, C.~Sab\'{i}n-Sanjuli\'{a}n\inst{5,6}, S.~J.~Smartt\inst{20}, V.~E.~Stroud\inst{16,28},
J.~Th.~van~Loon\inst{29}, J.~S.~Vink\inst{10} \and N.~R.~Walborn\inst{19}}

\offprints{C.~J.~Evans at chris.evans@stfc.ac.uk}

\authorrunning{C.~J.~Evans et al.}
   
\titlerunning{VLT-FLAMES Tarantula Survey}
   
\institute{UK Astronomy Technology Centre, 
           Royal Observatory Edinburgh, 
           Blackford Hill, Edinburgh, EH9 3HJ, UK
             \and 
           Scottish Universities Physics Alliance (SUPA),
           Institute for Astronomy,
           University of Edinburgh, 
           Royal Observatory Edinburgh, 
           Blackford Hill, Edinburgh, EH9 3HJ, UK
             \and 
           Astronomical Institute Anton Pannekoek, 
           University of Amsterdam, 
           Kruislaan 403, 
           1098 SJ, Amsterdam, The Netherlands
             \and
           Astronomical Institute, 
           Utrecht University, 
           Princetonplein 5, 3584CC, 
           Utrecht, The Netherlands
             \and
           Instituto de Astrof\'isica de Canarias, 
           E-38200 La Laguna, Tenerife, Spain 
             \and
          Departamento de Astrof\'isica, Universidad de La Laguna, 
          E-38205 La Laguna, Tenerife, Spain
             \and 
           European Southern Observatory, 
           Alonso de Cordova 1307, 
           Casilla, 19001, 
           Santiago 19, Chile
             \and
           Excellence Cluster Universe, 
           Boltzmannstr. 2, 85748 Garching, Germany
             \and
           School of Physics, University of Exeter, 
           Stocker Road, Exeter EX4 4QL, UK
             \and
           Armagh Observatory, College Hill, 
           Armagh, BT61 9DG, Northern Ireland, UK
             \and
           Institute of Astronomy \& Astrophysics, 
           National Observatory of Athens,
           I. Metaxa \& Vas. Pavlou Street, P. Penteli 15236, Greece
             \and
           European Southern Observatory, 
           Karl-Schwarzschild-Strasse 2, D87548, Garching bei Munchen, Germany
             \and
           Harvard-Smithsonian CfA, 
           60 Garden Street, Cambridge, MA 02138, USA
             \and
           University of Vienna, Department of Astronomy, 
           T\"{u}rkenschanzstr. 17, 1180, Vienna, Austria
             \and       
           Argelander-Institut f\"ur Astronomie der Universit\"at Bonn, 
           Auf dem H\"ugel 71, 
           53121 Bonn, 
           Germany
             \and
           Department of Physics and Astronomy, 
           The Open University, Walton Hall, Milton Keynes, MK7 6AA, UK
             \and
           Departamento de Astronomia, Universidad de Chile, 
           Camino el Observatorio 1515, Las Condes, Santiago, Chile
             \and
           Dept. of Physics \& Astronomy, 
           Hounsfield Road, 
           University of Sheffield, S3 7RH, UK
             \and
           Space Telescope Science Institute,
           3700 San Martin Drive,
           Baltimore,
           MD 21218,
           USA
             \and
           Department of Physics \& Astronomy,
           Queen's University Belfast,
           Belfast BT7 1NN, Northern Ireland, UK
             \and
           Institute of Astronomy, University of Cambridge,
           Madingley Road, Cambridge, CB3 0HA, UK
             \and
           Dept. of Physics \& Astronomy, 
           University College London, 
           Gower Street, 
           London, WC1E 6BT, UK
             \and
           ESA/STScI,
           3700 San Martin Drive, 
           Baltimore, MD 21218, USA
             \and
           Instituto de Astrof\'isica de Andaluc\'ia-CSIC, 
           Glorieta de la Astronom\'ia s/n, 
           E-18008 Granada, Spain
             \and
           Institute of Astronomy with NAO,
           Bulgarian Academy of Sciences, 
           PO Box 136, 4700 Smoljan, Bulgaria
             \and
           Centro de Astrobiolog\'ia (CSIC-INTA), 
           Ctra. de Torrej\'on a Ajalvir km-4, 
           E-28850 Torrej\'on de Ardoz, Madrid, Spain
             \and
           Universit\"ats-Sternwarte, 
           Scheinerstrasse 1, 81679 Munchen, Germany
             \and
           Faulkes Telescope Project, 
           University of Glamorgan, 
           Pontypridd, CF37 1DL, Wales, UK
             \and
           Astrophysics Group, 
           School of Physical \& Geographical Sciences, 
           Keele University, 
           Staffordshire, ST5 5BG, UK
}
\date{}

\abstract{The VLT-FLAMES Tarantula Survey (VFTS) is an ESO Large Programme that
has obtained multi-epoch optical spectroscopy of over 800 massive
stars in the 30~Doradus region of the Large Magellanic Cloud (LMC).
Here we introduce our scientific motivations and give an overview of
the survey targets, including optical and near-infrared photometry and
comprehensive details of the data reduction.  One of the principal
objectives was to detect massive binary systems via variations in
their radial velocities, thus shaping the multi-epoch observing
strategy. Spectral classifications are given for the massive
emission-line stars observed by the survey, including the discovery of
a new Wolf--Rayet star (VFTS~682, classified as WN5h), 2$'$ to the
northeast of R136.  To illustrate the diversity of objects encompassed
by the survey, we investigate the spectral properties of sixteen
targets identified by Gruendl \& Chu from {\em Spitzer} photometry as
candidate young stellar objects or stars with notable mid-infrared
excesses.  Detailed spectral classification and quantitative analysis
of the O- and B-type stars in the VFTS sample, paying particular
attention to the effects of rotational mixing and binarity, will be
presented in a series of future articles to address fundamental
questions in both stellar and cluster evolution.}

\maketitle 

\section{Introduction}\label{intro}

Massive stars have played an important role in galaxy evolution
throughout cosmological time via their intense winds, ultraviolet
radiation fields, chemical processing, and explosions.  
For instance, they dominate the rest-frame ultraviolet spectra of
star-forming, Lyman-break galaxies, in which multiple `starburst'
components (or mergers) are seen out to redshifts of at least
$z$\,$\sim$\,5 \citep{dbl10}.  Indeed, they are thought to have been a
major factor in the reionization of the early universe \citep{hl97},
comprising the dominant component in the earliest galaxies, in which
the star-formation rate appears to increase by a factor of ten over a
period of just 200\,Myr \citep{bouwens11}.

Population synthesis codes such as Starburst99 \citep[][]{l99,l10}
form the bridge between our understanding of the physics and evolution
of individual stars, and efforts to analyse entire populations on
galaxy scales via interpretation of their integrated light. The
sensitivity of the 8-10\,m class telescopes has enabled spectroscopy
of individual massive stars in galaxies at Mpc distances
\citep[e.g.,][]{araucaria,kud08}, including some in low-metallicity
dwarf galaxies, in which the local conditions are close to those found
in the early universe \citep[e.g.,][]{wlm,ic1613,e07}.  

Unfortunately, we are currently limited to observations of the
brightest stars in such galaxies.  Only in the Galaxy and Magellanic
Clouds can we potentially assemble large observational samples of
massive stars that span a wide range of metallicity and intrinsic
luminosities. Ultraviolet satellites have enabled substantial surveys
of terminal wind velocities \citep[e.g.,][]{hp89,pbh90} and rotational
velocities \citep{penny96,h97,pg09}.  However, both observational and
computational challenges have precluded the atmospheric analysis of a
large, coherent sample of O-type stars from optical spectroscopy, with
studies limited to several tens of objects at most
\citep[e.g.][]{herrero00,herrero02,rp04,mokiem05,rmsmc,rmlmc,mfast3}.
At lower masses, while large observational samples of early B-type
stars are available, recent results have highlighted some serious
problems regarding our understanding of massive-star evolution
\citep{ihapj,ih09,b11}.  This leaves us in a situation where
fundamental questions concerning the formation, evolution, and
eventual fate of massive stars remain unanswered, particularly in the
context of results that point to the majority of high-mass stars being
in binary/multiple systems \citep[e.g.][]{se10}.

The Tarantula Nebula (NGC\,2070, 30~Doradus -- hereafter `30~Dor') in
the Large Magellanic Cloud (LMC) is the brightest H\,{\sc ii} region
in the Local Group.  It is comprised of multiple generations of star
formation, with at least five distinct populations \citep{wb97},
similar to cluster complexes seen well beyond the Local Group
\citep[e.g.,][]{b06}.  At the heart of 30~Dor is Radcliffe~136
\citep[R136,][]{f60}, a massive, young stellar cluster 
\citep[1-2\,Myr; e.g.,][]{dk98,mh98} dominated by early O-type and hydrogen-rich
WN-type stars, some of which are thought to have current masses in
excess of 150\,M$_{\odot}$ \citep{c10}.  To the north and west of R136
there is significant molecular gas \citep{w78,j98} with embedded
massive stars, which has been suggested as a second generation of (perhaps
triggered) star formation around R136 \citep{wb97,w99,wmb02}.

By virtue of its location in the LMC, the distance to 30~Dor is well
constrained \citep[we adopt a distance modulus to the LMC of 18.5,
e.g.,][]{g00} and the foreground extinction is relatively low compared
to some of the most massive Galactic clusters.  Moreover, the
metallicity of the LMC ($\sim$50\% solar) is well matched to typical
results for high-redshift galaxies \citep[e.g.][]{erb06,nld08}.

Thus, 30~Dor provides an excellent opportunity to study a broad age
range of massive stars within a single complex of star formation. In
addition to the rich populations of O- and early B-type stars, there
are over twenty Wolf--Rayet (W--R) stars, \citep[including examples of
both WN and WC subclasses, e.g.,][]{bat99}, several transition Of/WN
stars (Crowther \& Walborn, in preparation), and candidates for
high-mass young stellar objects (YSOs) identified from mid-infrared
observations with the {\em Spitzer Space Telescope} (see
Section~\ref{yso}).  With its rich stellar populations, 30~Dor is the ideal laboratory in
which to investigate important outstanding questions regarding the
physical properties, binary fraction, chemical enrichment, and
evolution of the most massive stars, as well as the relation of
massive stars to the formation and evolution of stellar clusters
\cite[e.g.,][]{dksw,pmg10}. 

Building on experience from the VLT-FLAMES Survey of Massive Stars
\citep{e05,f2}, we introduce the VLT-FLAMES Tarantula Survey (VFTS), a
multi-epoch spectral survey of over 800 massive stars in the 30~Dor
region, including $\sim$300 O-type stars.  In a series of articles,
the VFTS data will be used to investigate the properties of stellar
winds and rotational mixing in O-type stars, and to extend studies of
rotational mixing in B-type stars.  In particular, the surface
abundances of O-type stars are expected to be modified via the effects
of rotational mixing but, to date, nitrogen abundances have not been
determined for a large observational sample.  The inclusion of new
models of N~\3 in the {\sc fastwind} model atmosphere code
\citep{p05,jjj}, combined with analytical methods such as those
developed by \citet{mokiem05} or the use of a large model grid, means
that such a critical test of evolutionary prodictions is now feasible.

30~Dor is known to have a rich binary population \citep{b09} --
a key feature of the VFTS is a multi-epoch observational strategy
to obtain clear indications of binarity\footnote{Hereafter we use
`binary' as shorthand for both true binaries and multiple systems.}
in the large majority of the targets.  This will add a valuable
component to quantitative analysis of the spectra and interpretation
of the results, enabling tests of the predictions of evolutionary
models that include all of the relevant physical processes for both
single stars and binary systems \citep[e.g.][]{lcy08}.  
The survey also includes multi-epoch spectroscopy in the inner part of
30~Dor with the FLAMES--ARGUS integral-field unit (IFU) to obtain an
improved estimate of the velocity dispersion of single stars in R136, which will
be used to determine the dynamical mass of the cluster.

This introductory article presents the significant observational
material in one resource, comprising details of target selection,
observational strategy and data reduction
(Section~\ref{target_overview}), optical and infrared photometry of
the targets (Section~\ref{phot}), spectral classifications for the
observed massive emission-line and cooler-type stars
(Section~\ref{class}), and a discussion of the spectral properties of
stars with published mid-infrared excesses (Section~\ref{yso});
concluding remarks are given in Section~\ref{summary}.

\section{Spectroscopy}\label{target_overview}

The Tarantula Survey employs three modes of the Fibre Large Array
Multi-Element Spectrograph \citep[FLAMES;][]{flames} instrument on the
VLT:
\begin{itemize}
\item{{\it Medusa--Giraffe:} The majority of the spectra were obtained using the Medusa
fibre-feed to the Giraffe spectrograph.  There are a total of 132
Medusa fibres available for science (or sky) observations, deployable
within a 25$'$ diameter field-of-view and with a diameter of 1\farcs2
on the sky.}
\item{{\it ARGUS--Giraffe:} The ARGUS IFU was
    used to observe five pointings of R136 and its environs in the
    central part of 30~Dor.  The IFU was used in the 0\farcs52
    per spatial pixel (spaxel) mode, such that its 22\,$\times$\,14
    microlenses provide a total field-of-view for a single pointing of
    12$''$\,$\times$\,7$''$ on the sky.}
\item{{\it UVES:} In parallel to the ARGUS observations, the fibre feed to the red
    arm of the Ultraviolet and Visual Echelle Spectrograph (UVES) was
    used to observe a small sample of stars in the inner part of
    30~Dor at greater spectral resolving power than that delivered by
    Giraffe.}
\end{itemize}

The target selection is discussed in Section~\ref{target_selection},
followed by a description of the reductions of the three spectroscopic
components of the survey (Sections~\ref{medusa}, \ref{argus}, and
\ref{uves}).

\subsection{Target selection}\label{target_selection}
The Medusa fibre configurations were prepared using two sources.  Within a
radius of $\sim$60$^{\prime\prime}$ of the core, targets were (with a
couple of exceptions) taken from Brian Skiff's reworking of the
astrometry from the $UBV$ catalogue of
\citet{s99}\footnote{ftp://cdsarc.u-strasbg.fr/pub/cats/J/A+A/341/98}.
Targets beyond this region were selected from preliminary reductions
of $B$- and $V$-band observations (with the B/123 and V/89 filters,
respectively) with the Wide-Field Imager (WFI) at the 2.2\,m
Max-Planck-Gesellschaft (MPG)/ESO telescope at La Silla (from
observations by J.~Alves, B.~Vandame \& Y.~Beletsky; programme
076.C-0888).  Four overlapping WFI fields were observed that cover
approximately one square degree -- our FLAMES observations primarily
include sources from the first (northeast) WFI pointing, supplemented
to the west by targets from the second.  The median seeing in the
reduced images is $\sim$0\farcs8, although the conditions were
non-photometric and stars brighter than approximately 14th magnitude
were saturated.

Source detections in the WFI frames were performed using the {\sc
daophot} package \citep{daophot} within {\sc iraf}.
The astrometry was calibrated using stars from the UCAC-2 catalogue
\citep{ucac2}, which has an astrometric precision to better than
0\farcs1 (sufficient for the diameter of the Medusa fibres).

In a bid to avoid significant selection biases we did not employ any
colour cuts on our input target lists. The only constraint imposed was
a magnitude cut (before final photometric calibration) of
$V$\,$\le$\,17\,mag, to ensure adequate signal-to-noise (S/N) in the
resulting spectra.  The Fibre Positioner Observation Support Software
(FPOSS) was used with our combined `Selman--Skiff' and WFI catalogue
to create nine Medusa configurations, hereafter referred to as fields
`A' through to `I'.  Exactly 1000 targets were observed with the
Medusa--Giraffe mode in these nine configurations. The reductions and
characteristics of these data are discussed below.  After rejection of
foreground stars (and a small number of others due to issues relating
to data quality, see Section~\ref{crosscont}), we have Medusa (or UVES)
observations of 893 unique targets.

A full catalogue of the survey targets is given in Table~\ref{cat} (published
online) in which they are RA-sorted and given a running number, which hereafter
serves as an identifier of the form VFTS~{\#\#\#}.  Given its
importance and rich stellar populations, 30~Dor has been the subject
of numerous studies over the past decades -- the final column of
Table~\ref{cat} provides a thorough (but not exhaustive) list of
previous aliases/identifications for our targets.

\begin{figure*}
\begin{center}
\includegraphics[scale=0.55]{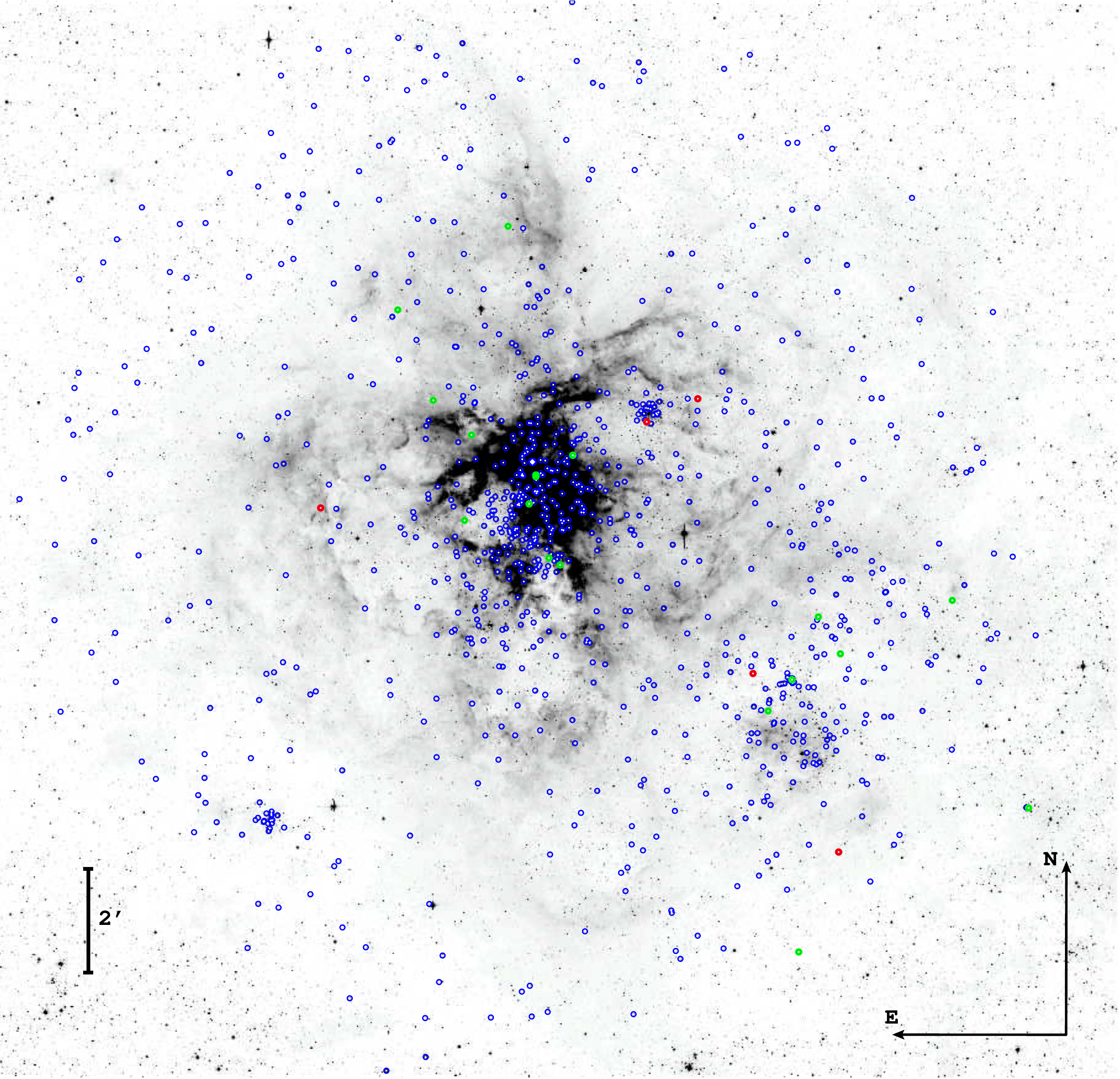}
\caption{FLAMES--Medusa targets overlaid on the $V$-band WFI image.  
The targets (blue circles) are primarily in the central part of 30~Dor, but also span
the broader region including Hodge~301 ($\sim$3$'$ NW of R136),
NGC\,2060 ($\sim$6\farcm5 SW), and SL~639 ($\sim$7\farcm5 SE).  To
highlight their locations, the emission-line stars discussed in
Section~\ref{wrstars} are encircled in green, and the five luminous
red supergiants from Section~\ref{coolstars} are encircled in
red.}\label{targets}
\end{center}
\end{figure*}

FLAMES has a corrected field-of-view of 25$^\prime$. This means that
with one central telescope pointing we were able to observe stars in
some of the local environs of 30~Dor as well as in the main body of
the H~\2 region.  The distribution of the Medusa (and UVES) targets is
shown in Figure~\ref{targets}, overlaid on a section of the $V$-band
WFI mosaic.  The targets are primarily located in the main 30~Dor
nebula (NGC\,2070), but also include at least three separate
associations: 1) Hodge~301, approximately 3$'$ to the northwest of
R136 \citep{h88}, which is somewhat older than the rest of 30~Dor
\citep[20 to 25\,Myr,][]{gc00}; 2) NGC\,2060 ({\it aka} 30~Dor~B and
LHA~120-N~157B), 6\farcm5 to the southwest of R136, which is
associated with a supernova remnant
\citep[e.g.][]{cksl92}; 3) SL~639 \citep{sl63}, 7\farcm5 to the
southeast.  The survey targets span a total diameter just in excess of
20$'$, with over 500 of them within a 5$'$ ($\approx$75\,pc) radius
from the centre of R136.  Distances to each target from R136
(specifically, {R136-a1}: $\alpha$\,$=$\,5$^{\rm h}$\,38$^{\rm
m}$\,42{\mbox{\ensuremath{.\!\!^{\,\rm s}}}}39,
$\delta$\,$=$\,$-$69$^{\circ}$\,06$'$\,02\farcs91, J2000.0) are given
in the fifth and sixth columns of Table~\ref{cat}, in arcmin and pc,
respectively.

\subsection{VLT Medusa--Giraffe spectroscopy}\label{medusa}

\subsubsection{Observational strategy}

The nine Medusa fibre configurations (`A'--`I') were observed using three
of the standard Medusa--Giraffe settings (LR02, LR03, and HR15N). These
provide intermediate-resolution spectroscopy of the lines commonly
used in quantitative analysis of massive stars in the
3960--5070\,\AA\/ region, combined with higher-resolution observations
of the H$\alpha$ line to provide a diagnostic of the stellar wind
intensity.

The wavelength coverage and spectral resolution ($\Delta$\lam, as
determined by the mean full-width at half-maximum of arc lines in the
wavelength calibration exposures) is summarised in
Table~\ref{arc_fwhm}.  We also give the effective resolving power, $R$
(i.e., \lam/$\Delta$\lam), for the central wavelength of each setting.

Most of the Medusa observations were obtained over the period 2008
October to 2009 February, with a seeing constraint of $\le$\,1\farcs2.
Pairs of exposures were taken in each observing block (OB).  To obtain
sufficient S/N for quantitative analysis of each target, three OBs at
both the LR02 and LR03 settings were observed, and two at HR15N.  No
explicit time constraints were placed on the execution of these OBs
and, in the majority of cases, the sequences at a given wavelength
setting were observed consecutively (but not in all, e.g., the LR03
observations of Field~D). The modified Julian Dates (MJD) for each OB
are given in Tables~\ref{mjd_medusa1} and \ref{mjd_medusa2} in the
appendix (published online).

The key observational feature of the survey is three repeat OBs at the
LR02 setting to detect radial velocity variables (epochs 4, 5, and 6
in the appendix). These were scheduled in the service queue such that
a minimum of 28 days had passed between execution of the third and
fourth epochs, similarly between the fourth and fifth.  The sixth
epoch was obtained in 2009 October.
The inclusion of the extra epoch a year later significantly helps with
the detection of both intermediate- and long-period binaries.  An
important part of the interpretation of these
multi-epoch spectra will be modelling of the detection sensitivities
to put firm limits on the observed binary fraction (Sana et al. in
preparation).

\begin{table}
  \caption{Summary of the wavelength coverage, exposure time
    per observing block (OB), and measured spectral
    resolution ($\Delta$\lam) of the different FLAMES modes and settings used in the survey.}\label{arc_fwhm}
\begin{center}
\begin{tabular}{llcccc}
\hline\hline
Mode & Setting & Exp. time/OB & \lam-coverage & $\Delta$\lam & $R$ \\
& & [s] & [\AA] & [\AA] & \\
\hline
Medusa & LR02 & 2$\times$1815 & 3960--4564 & 0.61 & \oo7\,000 \\
Medusa & LR03 & 2$\times$1815 & 4499--5071 & 0.56 & \oo8\,500 \\
Medusa & HR15N & 2$\times$2265 & 6442--6817 & 0.41 & 16\,000 \\
ARGUS  & LR02 & 2$\times$1815 & 3960--4570 & 0.40 & 10\,500 \\
UVES & 520 & 2$\times$1815 & 4175--6200 & 0.10 & 53\,000 \\
\hline
\end{tabular}
\tablefoot{The UVES data do not include the 5155 to 5240\,\AA\/ region because of
the gap between the two detectors.}
\end{center}
\end{table}

Owing to the nature of service-mode observations, a small number of the
OBs at the LR02 wavelength setting were repeated for operational
reasons, e.g., deterioration of the seeing beyond the required
constraints during the exposure, failure of the tracking on
the guide star, etc.  In cases where full exposures (i.e. 1815\,s)
were completed, these observations were retained -- even if the S/N is
lower than in other frames, there will be useful radial velocity
information for some targets.  This leads to the `extra' frames for
Fields~A, B, E, and I in Tables~\ref{mjd_medusa1} and
\ref{mjd_medusa2}.  There is also an extra LR03 exposure of Field~F
(`LR03~01\,[c]'), included out-of-sequence of its execution.  This OB
was aborted mid-execution because the telescope was being used for
interferometric observations -- the one completed exposure was
retained because it provides additional radial velocity information, for
example, in the case of R139 \citep{t11}.

\subsubsection{Reductions}\label{redux}
The ESO Common Pipeline Library (CPL) FLAMES reduction
routines (v2.8.7) were used for all of the initial data processing, i.e. bias
subtraction, fibre location and (summed) extractions, division by a
normalised flat-field, and wavelength calibration.  The subsequent
reduction stages were:
\begin{itemize}
\item{{\it Heliocentric correction:} All spectra were 
corrected to the heliocentric frame, using the {\sc iraf} packages
{\sc rvcorrect} and {\sc dopcor}.}
\item{{\it Sky subtraction:} 
Each sky fibre was inspected for signs of an on-sky source or
cross-contamination from bright spectra/emission lines from adjacent
fibres on the detector (see Section~\ref{crosscont}).  Any
contaminated sky fibres were rejected (typically one or two per frame)
brefore creating a median sky spectrum, which was then subtracted
from all fibres in that observation.}
\item{{\it Error spectrum:} An error spectrum was produced by the pipeline
for each fibre as part of the reduction process. For each wavelength
bin it recorded the statistical error arising from the different stages
in the reduction, e.g., bias level, detector gain, read-out noise.
These data were combined with the errors on the median sky spectrum to
obtain an estimate of the overall error for each spectrum.}
\item{{\it Cosmic-ray rejection:} Our Medusa--Giraffe exposures were taken in
    consecutive pairs within the OBs.  To clean the extracted spectra
    of cosmic rays we employed a technique developed (by I.D.H.) for
    the 2dF survey of the Small Magellanic Cloud (SMC) by \citet{eh04}.  For each spectrum, the
    ratio of the two exposures was calculated. A boxcar 4-sigma clip
    over 100 wavelength bins was then performed on this ratio value.
    Any unexpected and significant deviations in the ratio are
    indicative of a feature in only one of the observations. Because the
    exposures were consecutive, it is safe to assume that a
    transient feature was a cosmic ray. The pixels identified as
    suspect were rejected, then replaced with the value from the
    sister exposure, appropriately scaled by the ratio of the
    surrounding region.  This approach results in good removal of cosmics,
    but is not perfect (which would require inter-comparison between three exposures).}
\item{{\it Rejection of foreground stars:} Preliminary inspection of
    the LR02 data for each target was used to identify foreground
    cool-type stars to exclude them from our final catalogue (employing
    a velocity threshold of $v_{\rm r}$\,$<$\,100\,\kms\/ for
    rejection).  A total of 102 stars were excluded at this stage,
    which also included a small number of cool stars with LMC-like
    radial velocities but with very poor S/N.}
\end{itemize}

As an example of the final S/N of the spectra we consider VFTS~553, one
of the faintest Medusa targets ($V$\,$=$\,17.0\,mag, for which the
data reveals an early B-type spectrum).  The mean S/N ratio in the
individual exposures {\em per resolution element} were $\sim$50 for
all three wavelength settings (determined using line-free regions of
the stellar continuum). In contrast, the S/N for VFTS~527 (R139, one
of the brightest targets) exceeds 400 per resolution element in the best
spectra \citep{t11}.

\subsubsection{Nebular contamination}
Related to the sky subtraction, one of the principal limitations of
fibre spectroscopy is the subtraction of local nebular emission.
Indeed, at the distance of the Magellanic Clouds, even
(seeing-limited) long-slit spectroscopy can suffer difficulties from
spatially-varying nebular emission \citep[cf. the {\em HST} spectroscopy
from, e.g.][]{wmb02}.  Given the significant nebular emission in
30~Dor, the majority of the Giraffe spectra have some degree of
nebular contamination.  

The combination of extended nebular emission with the stellar flux
from a point-source means that, because the seeing conditions vary for
different epochs, the relative intensity of the nebular contamination
can vary.  Apart from a minority of spectra with particularly strong
emission, this is not a big problem for the hydrogen Balmer lines,
where the core nebular profile can (effectively) be ignored in
quantitative analysis \citep[e.g.][]{rmsmc}.  However, care is
required when selecting He~\1 lines for analysis (most notably
\lam4471) because small changes in the nebulosity could be interpreted as
evidence for binarity if unchecked with other lines.

\subsubsection{Cross-contamination of spectra}\label{crosscont}
To keep the observing as simple and homogeneous as possible, a
constant exposure time was used for each fibre configuration.  A
consequence was that bright stars can cross-contaminate adjacent
spectra of fainter stars on the detector after dispersion. 
This cross-contamination between fibres was generally minimal, but
inspection of the reduced spectra revealed a small number of fibres that
were contaminated by bright emission lines in W--R or `slash' stars and, 
in a couple of cases, by luminous supergiants with large continuum
fluxes.  

Thus, for the observation obtained in the best seeing, we inspected
the spectra adjacent to all `bright' stars (with counts greater than
10\,000) for each Medusa field and wavelength setting.  In some
instances the contaminated spectra were sky fibres (rejected as
described above), but five stellar targets were omitted because they
were pathologically contaminated by very strong emission lines.  A
further 22 stars have some element of cross-contamination -- typically
an artificial broad emission `bump' at \lam4686 from strong He~\2 in
emission-line stars in adjacent fibres.  These are noted in the final
column of Table~\ref{cat} -- quantitative spectral analysis of these
stars may not be possible, but the radial velocities from other
regions/lines will still be useful to investigate binarity, gas
dynamics, etc.

\subsection{VLT ARGUS--Giraffe spectroscopy}\label{argus} 

The Medusa data were complemented by five pointings within the central
arcminute of 30~Dor with the ARGUS IFU, as shown in the main panel of
Figure~\ref{fchart_argus}.  The coarser ARGUS sampling
was used (0\farcs52/microlens, giving a field-of-view of
11\farcs5$\times$7\farcs3), with a seeing constraint of $\le$\,0\farcs8.

These regions are densely populated with stars, particularly in the
core of R136. The first IFU pointing (`A1' in the figure) was located
on the core, with three pointings immediately adjacent.  The fifth pointing
was located to the NNE of R136 to target a reasonable number of stars
at a slightly larger radius from the core.

Full spectral coverage for quantitative analysis is best
obtained with AO-corrected or {\em HST} spectroscopy -- our intention
here was to probe the dynamics of these inner regions, again with
follow-up observations to identify binary systems.  Thus, only the
LR02 Giraffe setting was used.  The resulting wavelength coverage was
comparable to that from the Medusa observations, but at greater
resolving power because of the smaller aperture (see Table~\ref{arc_fwhm}).

Two OBs were observed without time restrictions and, similar to our strategy
for the Medusa data, follow-up epochs (third and fourth) were observed with
a minimum interval of 28 days.  All these data were obtained over
the period 2008 October to 2009 March, with a final (fifth) epoch
observed in 2009 December/2010 January.  

As with the Medusa observations, a number of the ARGUS OBs were
re-observed owing to changes in conditions and other operational issues.
The full list of completed exposures for the ARGUS frames is given in
Table~\ref{mjd_arg1}.  These epochs also apply to the UVES
observations (Section~\ref{uves}), which were taken in parallel to the
ARGUS data.

The seeing conditions are more critical to these IFU observations than
for the Medusa frames, so the adopted nomenclature is slightly
different for repeated OBs.  The exposures obtained under the best
conditions (ascertained from the seeing value recorded in the file
headers at the time of observation) are identified as the `a$+$b'
pair, with other exposures following in order of execution (as `c$+$d'
etc.), e.g., epoch four for the fourth pointing (`A4').

\subsubsection{Reduction}
The ARGUS frames were reduced using the same methods as for the Medusa
data apart from the sky subtraction and the combination of spectra
from adjacent spaxels on the sky (individual stars typically extend over
several spaxels).  The extracted (cosmic-clipped) spectra were
corrected to the heliocentric frame, then combined as follows.

Sources were selected if they appeared isolated in the reduced IFU
datacube and if they could be matched to a star (or multiple,
densely-packed stars) in an archival {\em HST} $F555W$ image (from the
early release science observations of 30~Dor taken with the Wide-Field
Camera Three, WFC3).  Less isolated sources were also extracted if
their spectra could be distinguished from those of their surroundings,
and if they had a matching bright source in the WFC3 image. (Some
sources are obviously multiple in the WFC3 image but their spectra
were retained because they might still prove useful in the analysis of
the velocity dispersion of R136.)  The spaxels combined for a given
source were selected on the basis that they showed the same spectral
features (and relative strengths) as the brighest/central spaxel of
that source.  There are small positional shifts of approximately one
spaxel between the different epochs of a given pointing.  These shifts
were taken into account when defining (for each frame) the spaxels
that contribute to each source.

\begin{center}
\begin{figure*}
\includegraphics[scale=0.25]{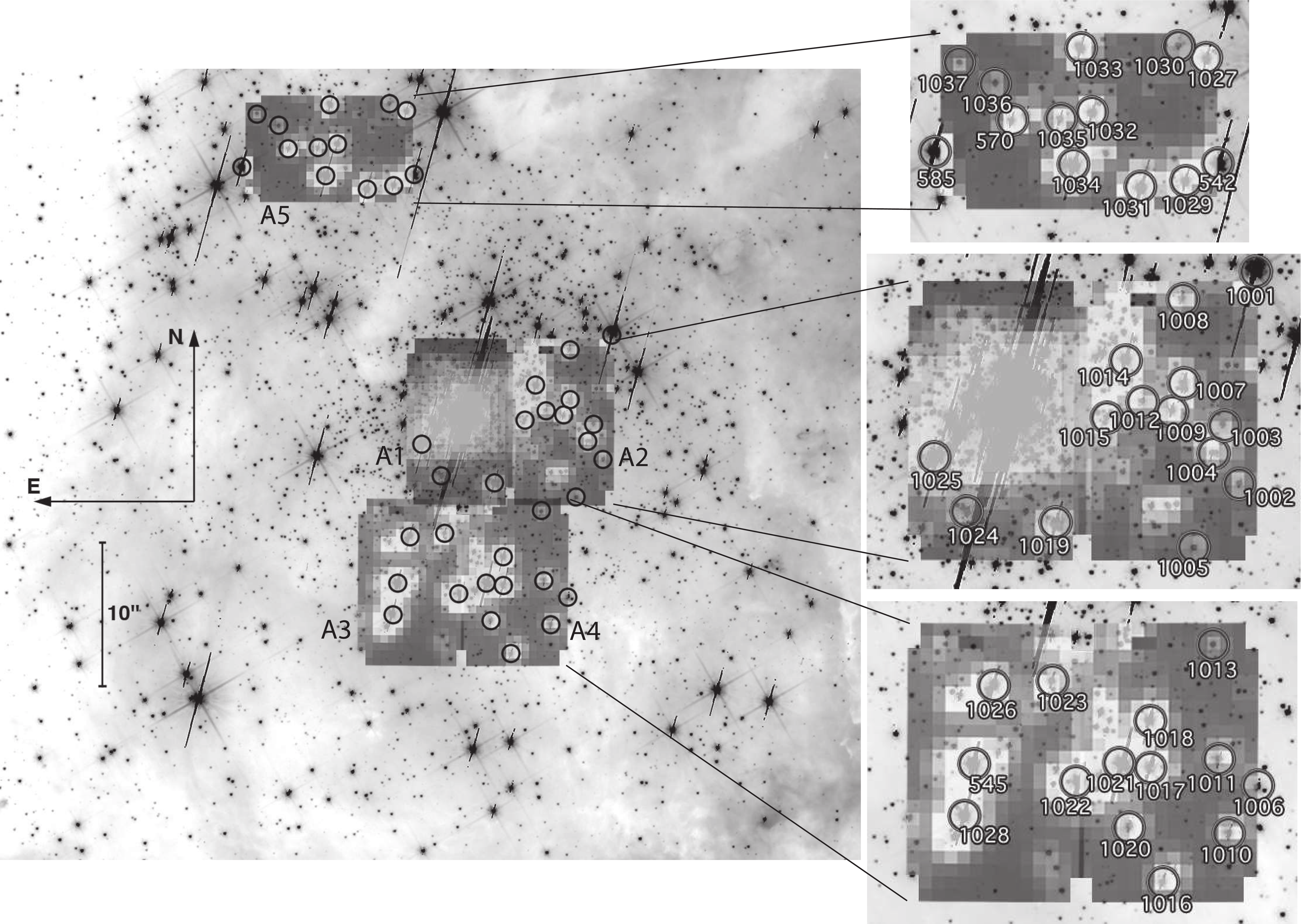}
\caption{{\em Main image:} the five ARGUS 
pointings overlaid on an F555W {\em HST}-WFC3 image. {\em
Right-hand panels:} identification of individual extracted sources --
{\it upper panel:} Field `A5'; {\it central panel:} Fields `A1' (eastern
pointing) and `A2'; {\it lower panel:} Fields `A3' (eastern pointing) and
`A4'.}\label{fchart_argus}
\end{figure*}
\end{center}

Spectra were extracted for a total of 41 sources from the ARGUS
frames.  The 37 unique ARGUS sources are appended to the end of
Table~\ref{cat}, with coordinates from centroids of matching sources
in the WFC3 image (transformed to the same astrometric system as the
Selman--Skiff catalogue for consistency).  To distinguish from targets
observed with Medusa and/or UVES, these sources are given a separate
series of RA-sorted identifiers, starting with VFTS~1001. Photometry
is available for all but three from \citet{s99}, as listed in
Table~\ref{cat}.  Note that four Medusa/UVES sources were also
observed with ARGUS (VFTS~542, 545, 570, and 585), as noted in the
final column of Table~\ref{cat}.  The five ARGUS pointings and the
location of the extracted sources are shown in
Figure~\ref{fchart_argus}.

The four or five spaxels with the lowest counts in each pointing were
used for local sky spectra. Even so, for most sources the nebular
subtraction is still far from perfect given the small-scale variations
discussed earlier.  Before combining the sky spectra, a 5$\sigma$-clip
(compared to the noise of the median spectrum) was employed to remove
remaining cosmic rays or artefacts.  A weighted-average sky spectrum
was created, then subtracted from the spectra from each of the source
spaxels.

Differential atmospheric refraction means that the apparent slope of
the continuum of a given source varies from spaxel to spaxel.  Thus,
each spectrum was normalised individually before combining the
contribution of different spaxels. This normalisation was done by a
spline-fit across carefully selected continuum regions, then division
of the spectrum by the resulting smooth curve.  This method generally
gave an excellent fit to the continuum, but is less certain for
spectra with broad emission lines, for which the continuum is hard to
define.  The final spectrum of each source is a weighted average of
the normalised spectra from the individual spaxels, again employing a
5$\sigma$-clip around the median to remove spurious pixels.

\subsection{VLT FLAMES--UVES spectroscopy}\label{uves}

The fibre feed to the red arm of UVES was used to observe 25 stars in
parallel to the ARGUS observations.  Twenty of these were also
observed with Giraffe -- our aim was to exploit the availability of
UVES to obtain broader spectral coverage and additional epochs for
radial velocity measurements (at even greater precision).  The UVES
targets are indicated by `U' in the second column of Table~\ref{cat},
and are all from the \citet{s99} catalogue, i.e., in the inner part of
30~Dor near R136.

\subsubsection{Reduction}

The two CCDs in the red arm of UVES were processed separately.  With
the \lam5200 central wavelength setting, the short-wavelength CCD
provided coverage of 4175 to 5155\,\AA, with the long-wavelength CCD
spanning 5240 to 6200\,\AA.  The UVES CPL routines (v4.3.0) from ESO were used
for the preliminary reduction stages: bias correction, fibre
extraction, wavelength calibration, and division by a normalised
flat-field frame.  Our own {\sc idl} routines were then used for
merging of the orders and velocity correction to the heliocentric
frame.  For these bright targets and at this resolving power, the sky
background was relatively low -- given the problems relating to merging
of the echelle orders, subtraction of the sky spectra resulted in a
poorer final data product, so they were not used.  From analysis of
the reduced arc calibration frames, the delivered resolving power was
$R$\,$\sim$\,53\,000.

\section{Photometry}\label{phot}

The approximate photometric calibration of the WFI imaging was
sufficient for target selection (Section~\ref{target_selection}), but
quantitative spectroscopic analysis requires precise
photometry.  Given the complexity of 30~Dor (in terms of source
detection in regions of bright nebulosity) and the non-photometric
conditions/saturation of bright stars in the WFI observations, we 
revisited the data with a more refined photometric analysis.

From combining the Selman--Skiff catalogue and our analysis of the WFI
frames we obtained $B$- and $V$-band photometry for 717 of the VFTS
targets.  These were then supplemented by photometry for 68 stars from
\citet{p93} and for 91 stars from new imaging at the Cerro Tololo
Inter-American Observatory (CTIO), Chile.  Each of these photometric
sources is discussed briefly below.

Moreover, when undertaking quantitative spectral analysis of massive
stars in regions of variable extinction, near-IR photometry can be a
useful input towards the determination of accurate stellar luminosities
\citep[e.g.][]{c10}. Cross-matches of the VFTS targets with published
near-IR data are discussed in Sections~\ref{nirphot} and \ref{2mass}.

We note that there is a significant amount of imaging of selected regions
of 30~Dor at better spatial resolution, principally from the {\em
HST}. Our philosophy here is to compile optical photometry from
seeing-limited, ground-based images, which are reasonably matched to
the aperture of the Medusa fibres. Thus, even if our `stars' are
actually multiple objects, the light received by the fibre is
comparable to that measured from the adopted imaging.

\subsection{`Selman' photometry}\label{phot_selman}

\begin{center}
\begin{figure}
\includegraphics[width=8.75cm]{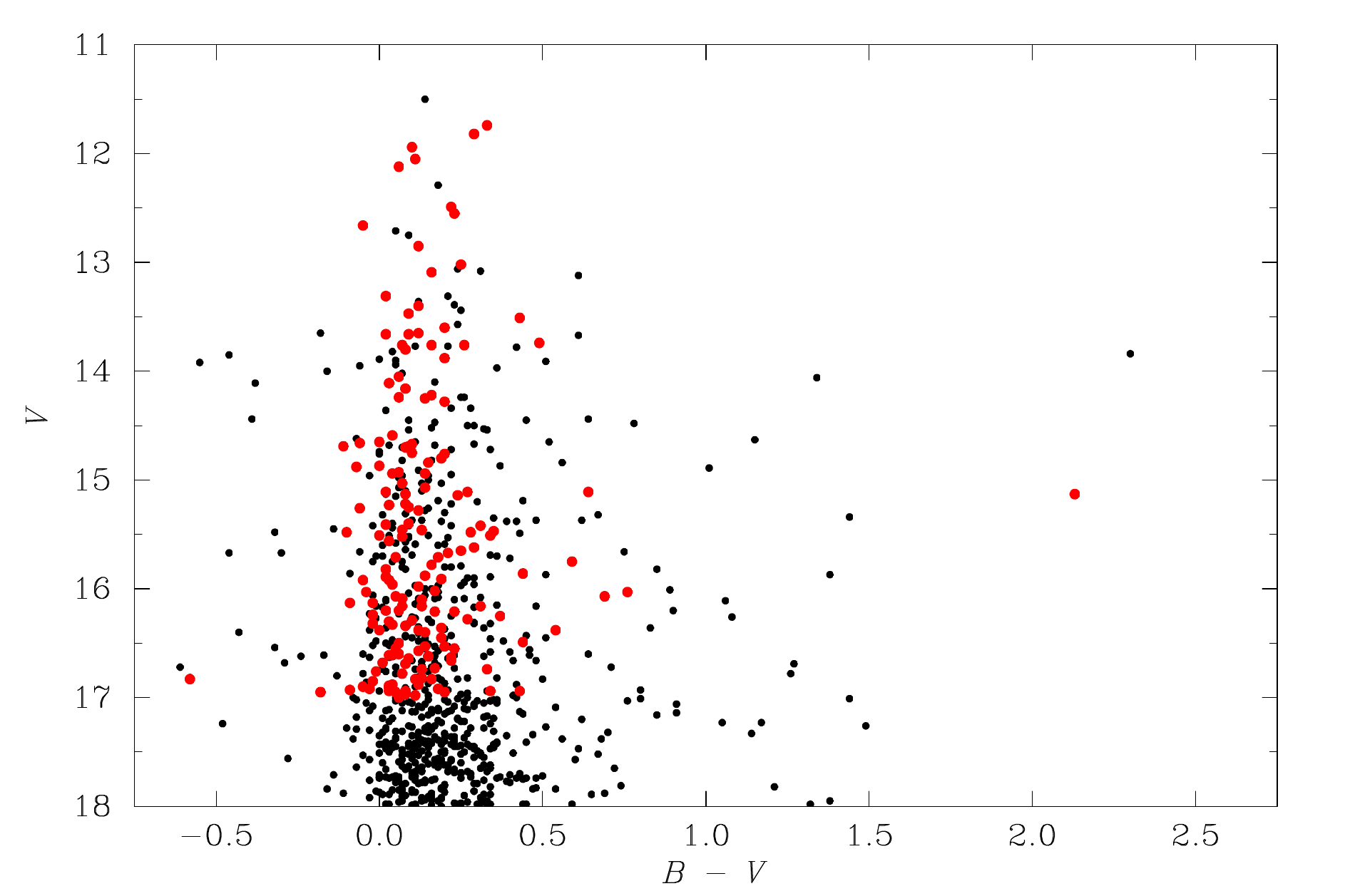}
\caption{FLAMES targets with photometry from \citet{s99} (167 stars, red points) compared to all
sources from their catalogue with $V$\,$\le$\,18 (black).}\label{cmd_selman}

\includegraphics[width=8.75cm]{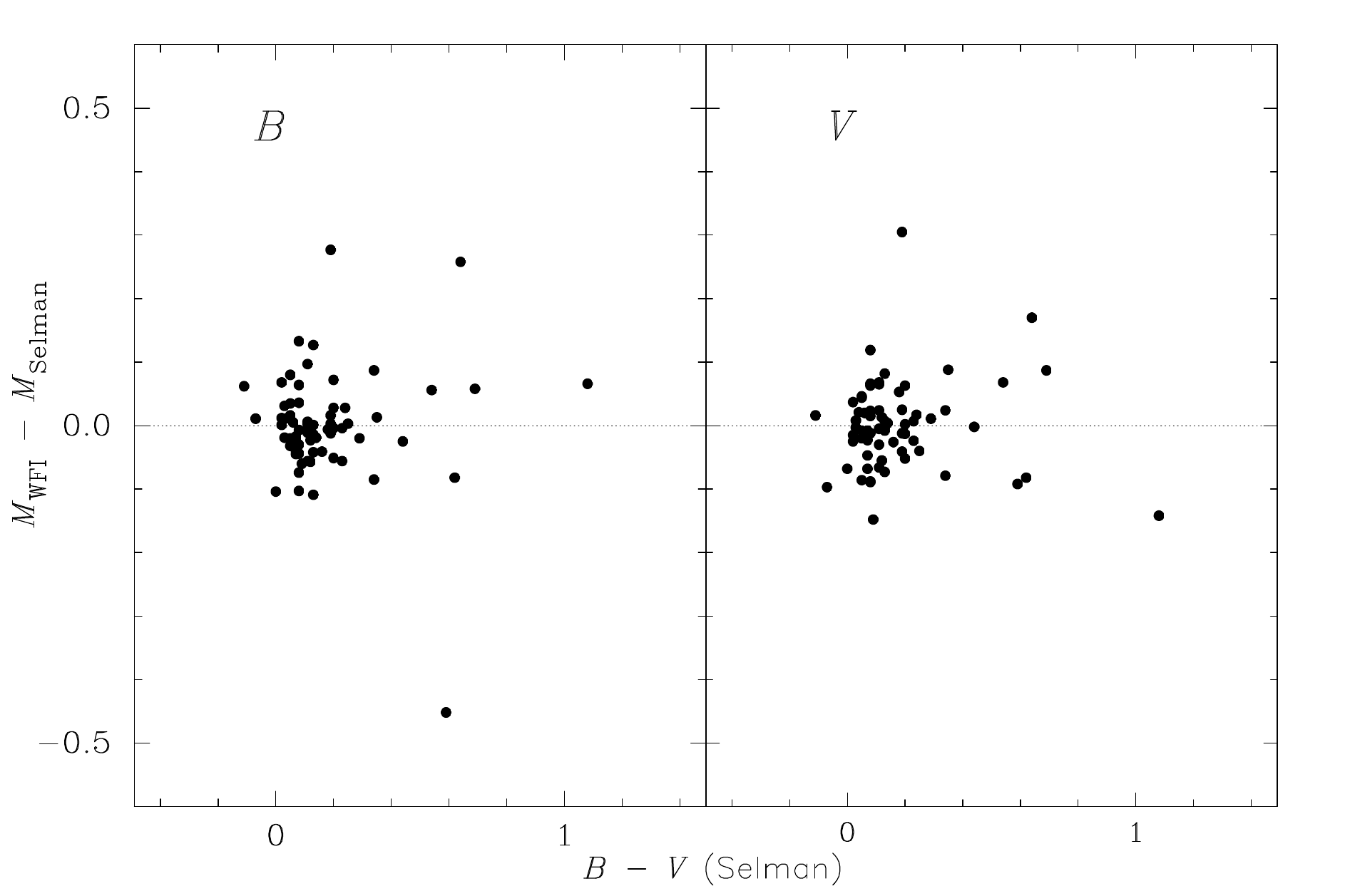}
\caption{Photometric residuals for the WFI data (where both are
WFI\,$-$\,Selman) as a function of $(B - V)_{\rm Selman}$.}\label{wfiZPs_BV}
\end{figure}
\end{center}

The data used for target selection in the central region were taken from the
Selman--Skiff catalogue, which was obtained using the Superb-Seeing Imager
(SUSI) on the 3.5\,m New Technology Telescope (NTT) at La Silla,
Chile. The conditions were photometric with sub-arcsecond seeing, and
short exposures (10--30s) were used to avoid saturation.  The fine
pixel scale of SUSI (0\farcs13/pixel) means that crowding is less
problematic than in the WFI frames.  We therefore adopt the $B$- and
$V$-band photometry from \citet{s99} for the 167 stars observed from
their catalogue, listed with the reference `S' in Table~\ref{cat}.  To
illustrate the selection effects employed on the FLAMES sources (i.e.
$V$\,$\le$\,17\,mag and no colour constraint), a colour--magnitude
diagram for the 167 stars with photometry from \citeauthor{s99} is
shown in Figure~\ref{cmd_selman}; their other sources
($V$\,$\le$\,18\,mag) are also shown.

\subsection{WFI photometry}\label{phot_wfi}

Our primary source of photometry is from point-spread function (PSF)
fitting of sources in the WFI frames with the stand-alone version of
{\sc daophot}. We are mainly interested in the relatively bright
population so we adopted a detection
threshold of 20 times the root-mean-square noise of the
local background to avoid spurious detections.  We then applied a cut of
0.2\,$<$\,sharpness\,$<$\,1 to further filter those detections for
extended objects, etc.  The resulting $B$- and $V$-band catalogues
were merged together using the {\sc daomatch} and {\sc daomaster}
routines \citep{daomatch}, i.e. the final catalogue includes only
those sources detected in both bands.

Zero-points for the WFI photometry were determined with reference to
67 stars (within a search radius of less than 0\farcs4 and
$V$\,$\le$\,18.0\,mag) that overlap with the Selman--Skiff catalogue.
The resulting residuals for these matched sources are shown in
Figure~\ref{wfiZPs_BV}, with standard deviations of $\le$0.1\,mag in
both bands. The second WFI pointing was bootstrapped from the first
using the mean differences between overlapping stars.

Photometry of 550 stars is adopted from analysis of the WFI images
\citep[calibrated using the catalogue from][]{s99}.
These are listed with the reference `W' in Table~\ref{cat}.
Their colour--magnitude diagram is shown in Figure~\ref{cmd_wfi};
also shown are the $\sim$2000 other WFI sources ($V$\,$\le$\,18\,mag)
within a radius of 12$'$ from R136.  There are ten sources with WFI
photometry in Table~\ref{cat} with $V$\,$\le$\,14.5\,mag, each of
which agrees with the CTIO photometry (Section~\ref{ctio}) to within
0.1\,mag, i.e., within the dispersion of the calibrations in
Figure~\ref{wfiZPs_BV}.

\subsection{`Parker' photometry}\label{parker}

\begin{center}
\begin{figure}
\includegraphics[width=8.75cm]{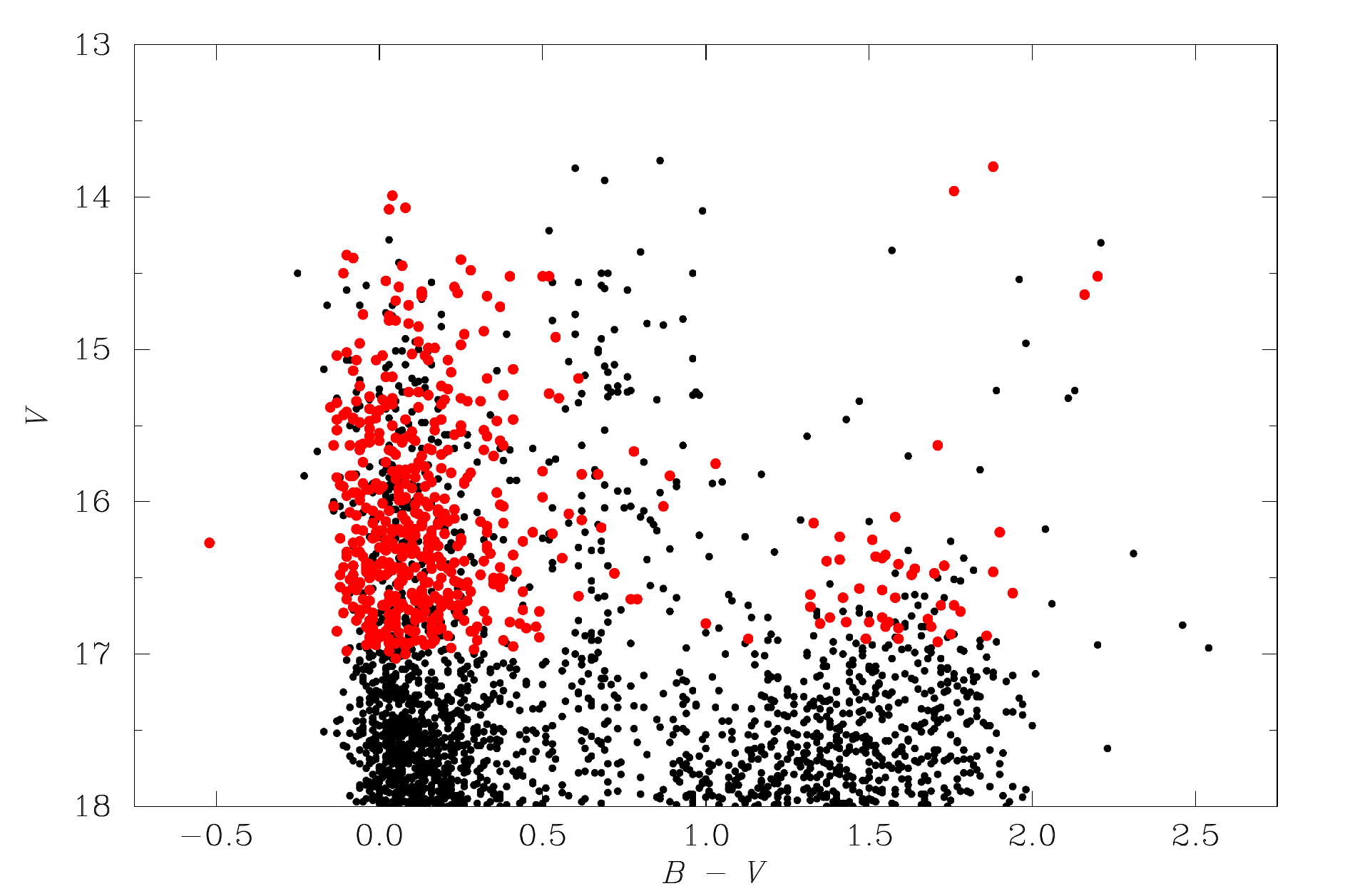}
\caption{FLAMES targets with WFI photometry (550 stars, red points) compared to all
WFI sources with $V$\,$\le$\,18 within a 12$'$ radius of the field centre (black).}\label{cmd_wfi}

\includegraphics[width=8.75cm]{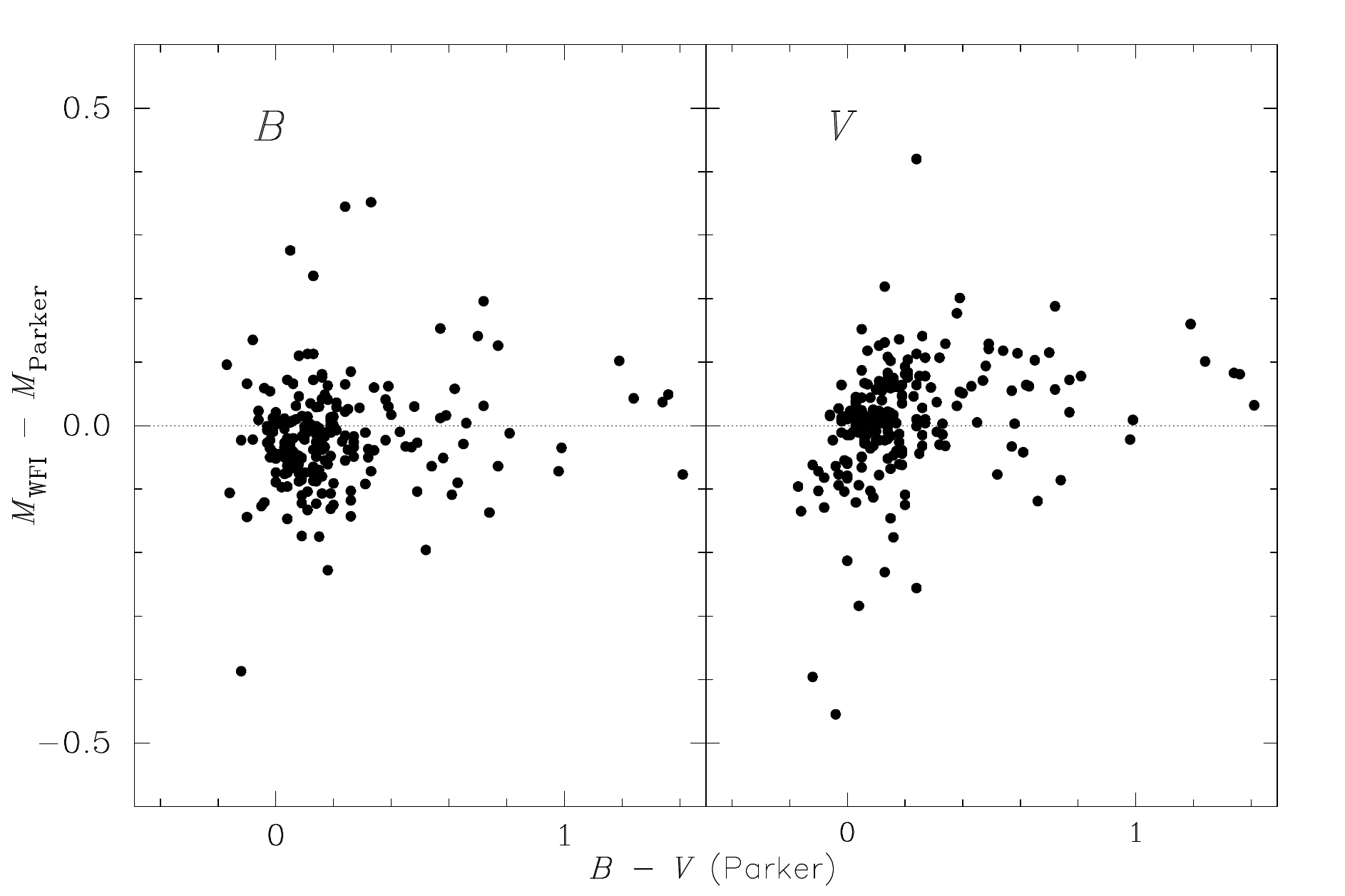}
\caption{Photometric residuals for the calibrated WFI data compared to those from 
\citet{p93} as a function of $(B - V)_{\rm Parker}$.}\label{wfi_x_parker}
\end{figure}
\end{center}

To supplement the WFI and Selman photometry we first turned to the
catalogue from \citet{p93} in 30~Dor, using the reworked astrometry from
Brian Skiff\footnote{ftp://cdsarc.u-strasbg.fr/pub/cats/II/187A}.  A
comparison between the Parker results and the calibrated WFI photometry for
219 matched stars (within a radius of 0\farcs5) yielded only small
residuals of $\Delta V$\,$=$\,0.01 ($\sigma$\,$=$\,0.2) and $\Delta
B$\,$=$\,$-$0.03 ($\sigma$\,$=$\,0.2)\,mag, as shown in
Figure~\ref{wfi_x_parker}.

A radial search (of 0\farcs5) and then visual matching of the VFTS
targets without WFI or Selman photometry yielded 68 matches in the
Parker catalogue (with a median radial offset of 0\farcs11).  For our
current purposes, we adopt his photometry for these 68 stars, listed
with the reference `P' in Table~\ref{cat}.

\subsection{MCPS photometry}\label{phot_mcps}

As a potential source of photometric information for the remaining 108
stars we turned to the LMC catalogue from the Magellanic Clouds
Photometric Survey \citep[MCPS;][]{mcps}, which includes photometry of
some of the bright sources from \citet{m02}.  However, we found large 
uncertainties on the MCPS photometry in the 30~Dor region.

\begin{center}
\begin{figure}
\includegraphics[width=8.75cm]{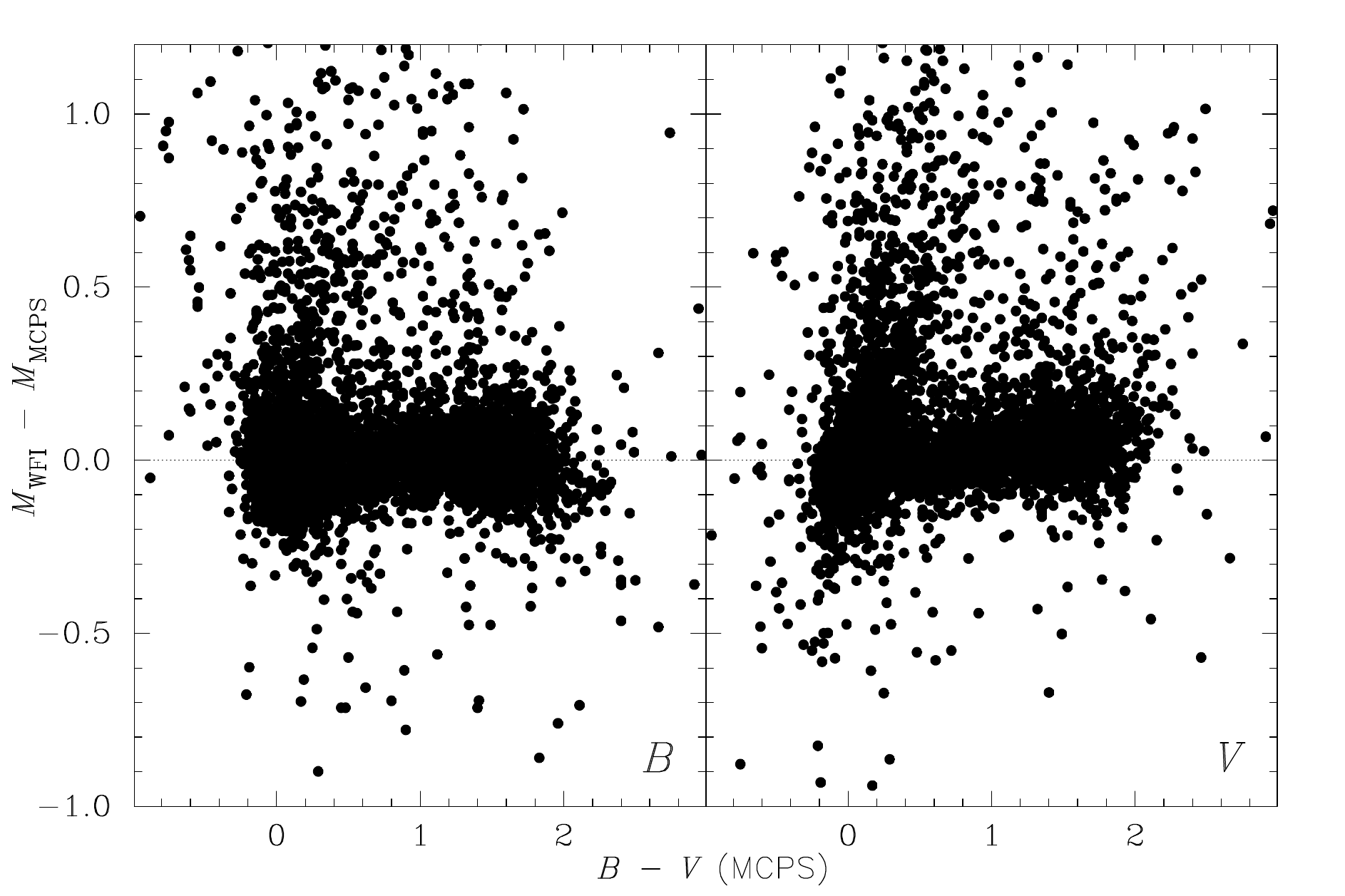}
\caption{Photometric residuals for the WFI data (where both are 
WFI\,$-$\,MCPS) as a function of $(B - V)_{\rm
MCPS}$.}\label{wfi_x_mcps}
\end{figure}
\end{center}

We first considered cross-matched sources from the MCPS catalogue
as an external check on the photometric calibration of the WFI frames.
Adopting a search radius of less than 1\farcs0, a total of 5726
`matches' were found between the northeastern WFI frame and the MCPS.
The residuals for $B$ and $V$ (in the sense of WFI\,$-$\,MCPS) are
shown in Figure~\ref{wfi_x_mcps} as a function of MCPS colour.  The
standard deviation of the residuals in both bands is $\sim$0.25\,mag,
not unreasonable considering the potential for specious matches and
complications relating to blending and nebulosity in this field --
the typical seeing from the MCPS is 1\farcs5 (with 0\farcs7 pixels),
but extended up to 2\farcs5 \citep{mcps}, while the photometry from
\citet{m02} used an aperture of 16\farcs2.  Although there is
reasonable agreement in the zero-point calibrations between the WFI
and MCPS data, the `plume' to brighter MCPS magnitudes is particularly
notable in Figure~\ref{wfi_x_mcps}.

\begin{center}
\begin{figure}
\includegraphics[width=8.75cm]{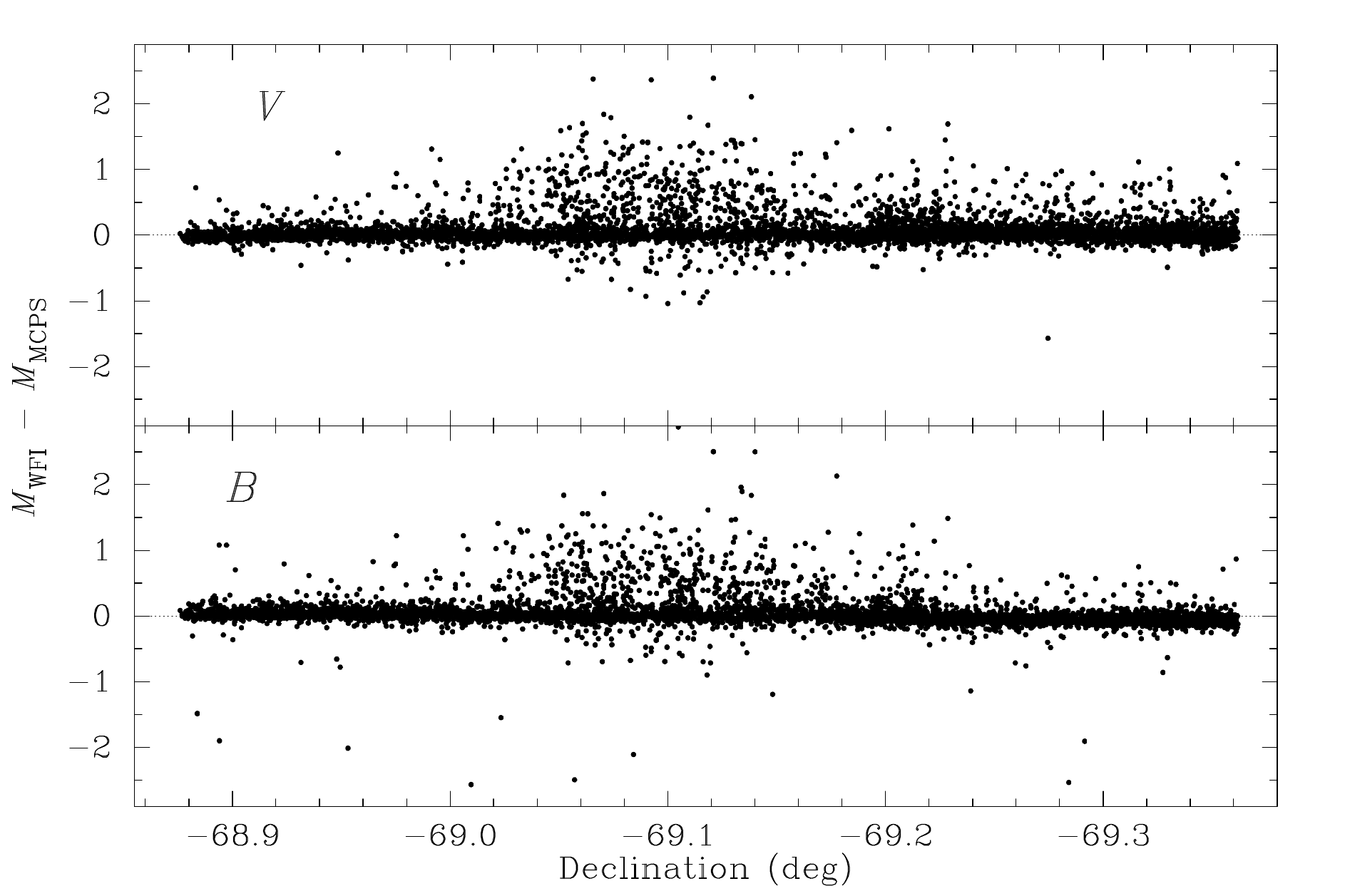}
\caption{Photometric residuals for the WFI data (where both are 
WFI\,$-$\,MCPS) as a function of declination.}\label{wfi_x_mcps_dec}

\includegraphics[width=8.75cm]{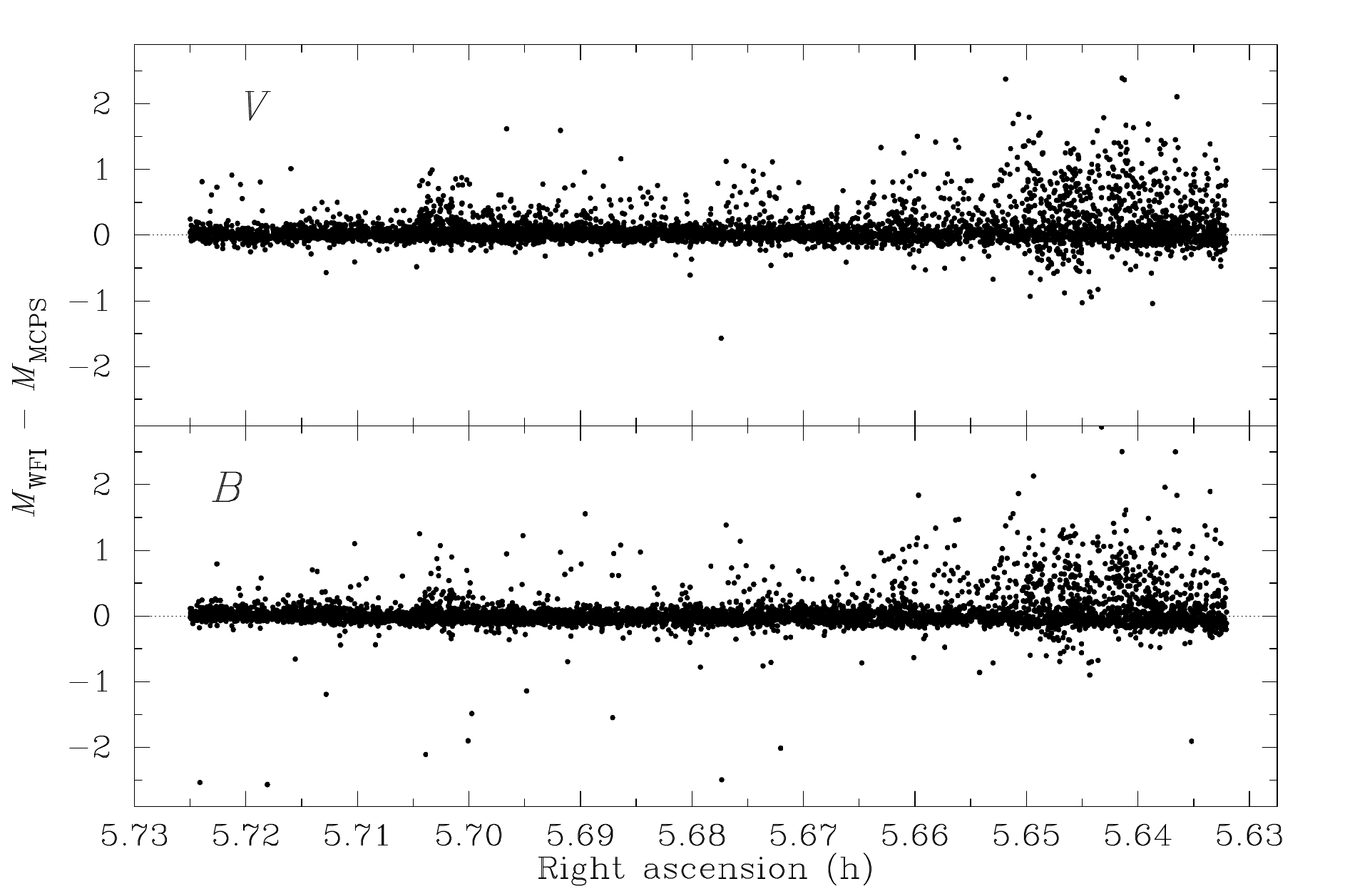}
\caption{Photometric residuals for the WFI data (where both are 
WFI\,$-$\,MCPS) as a function of right ascension.}\label{wfi_x_mcps_ra}
\end{figure}
\end{center}

Concerned by the potential of inter-CCD calibration problems in the WFI
data, we investigated the distribution of the residuals as a function
of both declination and right ascension, as shown in
Figures~\ref{wfi_x_mcps_dec} and \ref{wfi_x_mcps_ra}, respectively.
The residuals between the WFI and MCPS photometry are noticeably
larger over the main body of 30~Dor (centred at
$\alpha$\,$=$\,5{\mbox{\ensuremath{.\!\!^{\,\rm h}}}}645,
$\delta$\,$=$\,$-$69 {\mbox{\ensuremath{.\!\!^{\circ}}}}101, and with
the densest nebulosity having a diameter of $\sim$6$'$).  Larger
residuals can also be seen in the region of
$\alpha$\,$=$\,5{\mbox{\ensuremath{.\!\!^{\,\rm h}}}}702, the
centre of the dense cluster NGC\,2100.  The residuals are
predominantly in the sense of brighter magnitudes from MCPS compared
to the WFI frames, which suggests that they primarily arise from
unresolved blends or nebular contamination.  

\subsection{CTIO photometry}\label{ctio}

Thus, to obtain optical photometry for the majority of the remaining FLAMES
targets, we observed 30~Dor on 2010 December~27 with the Y4KCAM camera
on the CTIO 1\,m telescope, operated by the SMARTS
consortium\footnote{http://http://www.astro.yale.edu/smarts}. The
camera is equipped with an STA 4064\,$\times$\,4064 CCD
with 15\,$\mu$m pixels, yielding a scale of
0\farcs29\,pixel$^{-1}$ and total field-of-view of
20$'$\,$\times$\,20$'$ at the Cassegrain focus. 

Observations were obtained in photometric conditions with good seeing
(less than 1\farcs3), using $B$- and $V$-band filters in the Kitt
Peak system\footnote{http://www.astronomy.ohio-state.edu/Y4KCam/filters.html}.
A range of exposure times in each band were used to avoid saturating
bright stars (from 10\,s up to a maximum of 400 and
300\,s in the $B$- and $V$-bands, respectively).
Instrumental magnitudes were obtained using PSF-fitting routines in
{\sc daophot}, then observations of standard stars taken on the same
night in Selected Area~98 \citep{landolt} were used to transform the
photometry to the Johnson--Kron--Cousins system and to correct for
atmospheric extinction.  The standard observations spanned a range of
airmass (from 1.05 to 2.60) and included stars over a broad range of
colours ($-$0.3\,$\le$\,$(B-V)$\,$\le$\,1.7\,mag).

Robust matches were found for 91 of the remaining FLAMES targets, with
photometry from the CTIO imaging listed in Table~\ref{cat} with the
reference `C'.  Given the independent calibration using published
standards, we do not attempt to transform the photometry of these
stars onto the same exact system as the WFI photometry.  However, they agree
reasonably well -- a comparison of nearly 800 matched stars yields
mean residuals of $\Delta V$ and $\Delta B$\,$\le$\,0.05\,mag (with
$\sigma$\,$\sim$\,0.2\,mag in both bands).

Seven of the remaining targets were beyond the western and southern
extent of the CTIO images (VFTS~002, 003, 014, 016, 017, 739, and 764)
and five did not have counterparts in the CTIO catalogue owing to nearby
blends or the PSF-fitting criteria (VFTS~092, 172, 301, 776, and 835).
Photometry is available from the MCPS catalogue for four of these
(VFTS~016, 092, 739, 764), each of which is over 8$'$ from the core of
30~Dor, so the problems discussed in Section~\ref{phot_mcps} should be
minimised.

The last five targets without optical photometry (VFTS~145, 147, 150,
151, \& 153) are in the dense `Brey 73 complex', which was first resolved by
\citet{tld88} and later observed with the {\em HST} by
\citet{w95,wd99}.  The {\em HST} photometry from \citet{wd99} is
adopted in Table~\ref{cat} for the three (visually) single stars:
VFTS~145, 150, \& 153. The two brightest members (\#1 and \#2 from
\citeauthor{tld88}, VFTS~147 and 151) were resolved into separate
bright components by the {\em HST} imaging, i.e. the Medusa fibres
will contain contributions from these, and any future analysis will
have to consider their relative fluxes (and colours).

\subsection{Near-IR photometry}\label{nirphot}

The extensive near-IR imaging survey of the
Magellanic Clouds from \cite{irsf}\footnote{Obtained using the
Simultaneous three-colour InfraRed Imager for Unbiased Survey (SIRIUS)
camera (0\farcs45 pixel$^{-1}$) on the
InfraRed Survey Facility (IRSF) 1.4\,m telescope at Sutherland, South
Africa.} provides $JHK_{\rm s}$ photometry for nearly all of our
FLAMES targets. The mean seeing at $J$, $H$, and $K_{\rm s}$ was
1\farcs3, 1\farcs2, and 1\farcs1, respectively, i.e., well-matched to
the on-sky Medusa fibre aperture.

To identify IRSF counterparts to the FLAMES targets we employed an
astrometric search radius of 0\farcs5, then overlaid the resulting
list on the WFI $V$-band images to reject specious matches. The IRSF
magnitudes ($JHK_{\rm s}$, and their associated photometric errors)
are given for the FLAMES targets in Table~\ref{irsfcat} (published
online).  The IRSF `quality' flag in the final column indicates the
source detection in each of the three bands as follows \citep[for
futher details see][]{irsf}: `1', point-like; `2', extended source;
`3', saturated; `4', faint; `5', odd shaped (e.g. double sources);
`0', no detection.

The Two Micron All Sky Survey \citep[2MASS;][]{2mass} catalogue was
used to calibrate the IRSF astrometry and to provide checks on the
photometry \citep[see][]{irsf}.  However, we note that the
IRSF--SIRIUS filter-set was designed to match that of the Mauna Kea
Observatories near-IR filters \citep{tsv02}, and is therefore slightly
different to that used by 2MASS.  Transformation equations between the
IRSF--SIRIUS and 2MASS systems are given by \citet{irsf} and
\citet{k08}; in practise, the corrections between the two are
relatively small, with typical differences of less than 0.05\,mag
for $(J-K_{\rm s})_{\rm IRSF}$\,$<$\,1.7\,mag \citep{k08}.

There are five FLAMES targets without good IRSF matches within 0\farcs5:
VFTS~275, 503, 620, 823, and 828.  Additionally, VFTS~151 has two
potential matches, but is excluded following the discussion of
multiplicity in Section~\ref{ctio}.  In some instances there
were two potential matches both within a radius of less than
0\farcs3.  In approximately half of these one set of the IRSF
measurements was obtained in the `periphery' of the dither patterns
(so the S/N is lower) and these values were omitted.  However, for seven
targets (VFTS~125, 330, 368, 374, 377, 383, 384) both IRSF detections
are good spatial matches and not saturated.  For these sources we
compared the observational conditions of the relevant exposures
\citep[Table~2,][]{irsf} and adopted the values obtained in the best
seeing.  As noted in Table~\ref{cat}, VFTS~240 appears slightly
extended (or as a blended source) in the optical WFI image; the IRSF
catalogue has a counterpart approximately 0\farcs5 from the FLAMES
position, with $J$\,$=$\,16.31\,$\pm$\,0.06\,mag, but it was not
detected in the other two near-IR bands.

Similar comparisons were undertaken between the IRSF catalogue and the
unique ARGUS targets (i.e. those numbered from 1001 to 1037, see
Section~\ref{argus}).  Good matches (all within 0\farcs3) were found
for 31 sources, as summarised in Table~\ref{irsfcat}.  Six ARGUS
sources are without IRSF photometry: VFTS~1012, 1014, 1015, 1019,
1024, and 1025.

\subsection{Cross-references with 2MASS}\label{2mass}

Near-IR photometry for some of our targets is also available from the
2MASS catalogue. However, with a pixel size of 2\farcs0, its spatial
resolution is inferior to the IRSF data.  Nonetheless, we include
cross-identifications to 2MASS for the VFTS targets in the final
column of Table~\ref{cat} for completeness.  Our approach was similar
to the comparison with the IRSF catalogue, with visual inspection of
potential matches to reject those notably blended in the WFI data or
with positions offset by contributions from other nearby
stars/nebulosity.  In particular, given the crowding in the central
region around R136 and the limited angular resolution of 2MASS, we did
not attempt cross-matching within the central 30$''$.  Only
cross-matches with 2MASS photometric qualities of either `A' or `B'
(i.e. S/N\,$\ge$\,7) were retained, leading to 2MASS identifications
for 227 of our targets.

\section{Spectral classification}\label{class}

Following reduction, the first epoch of LR02 and LR03 spectra of each
target were inspected; approximately 300 targets display He~\2
absorption, which is indicative of an O (or B0) spectral type.  However, the
multi-epoch nature of the VFTS spectroscopy complicates precise
classification and entails significant analysis, which is being
undertaken as part of studies towards binarity and the determination of
stellar radial velocities (Sana et al. and Dunstall et al., both in
preparation). Detailed classifications of the O- and B-type spectra
will therefore be given elsewhere.  Here we present classifications for the
two smaller spectral groups in the survey: the W--R/massive
emission-line stars and the cooler stars (of A-type and later).

\subsection{Wolf--Rayet and `slash' stars}\label{wrstars}

Massive emission-line stars in the LMC have been well
observed over the years, from the seminal Radcliffe Observatory study
by \citet{f60}, to narrow-band imaging surveys to identify W--R
stars \citep[e.g.][]{ab79,ab80}, comprehensive spectroscopic
catalogues \citep[e.g.][]{brey,bat99}, and monitoring campaigns to
study binarity \citep[e.g.][]{m89,schnurr08}.

The W--R and transitional `slash' stars (see Crowther
\& Walborn, in preparation) observed by the VFTS are summarised in 
Table~\ref{spec_emstar} (and are highlighted in green in
Figure~\ref{targets}).  In addition to those in R136 \citep{c10}, these
stars comprise some of the most massive stars in 30~Dor and will
provide valuable insights into some of the most critical phases of
massive-star evolution.

Previous spectral types are included in Table~\ref{spec_emstar}, with
revised classifications in the final column if new features and/or
evidence for companions are present in the FLAMES data.  
The new data reveal massive companions in at least three of these
objects: VFTS~402  and 509 (BAT99-95 and 103,
respectively).  Quantitative analysis of these data is now underway.
Note that VFTS~527 ({\it aka} R139), has been
treated as a W--R star by some past studies; the VFTS 
observations have revealed it as an evolved, massive binary system
comprised of two O~Iaf supergiants \citep{t11}.

\begin{table*}
  \caption{Summary of published and, where relevant, new spectral 
    classifications for massive emission-line stars 
    in the VLT-FLAMES Tarantula Survey (VFTS) observations.  
    The spectral types adopted for the survey
    are given in bold font.  Aliases to the
    \citet[BAT99,][]{bat99} and \citet{brey} catalogues are provided (other aliases
    are included in Table~\ref{cat}).}\label{spec_emstar}
\begin{center}
\begin{tabular}{cccll}
\hline\hline
\multicolumn{3}{c}{Identification} & & \\
VFTS & BAT99 & Brey & Published spectral types & VFTS spectral types \\
\hline
002 & 85 & 67 & WC4$+$OB [SSM90] & {\bf WC4$+$O6-6.5~III} \\ 
019 & 86 & 69 & WN4 [Brey]; WN3o$+$O9: [FMG03] & {\bf WN3o} \\
079 & 88 & \phantom{a}70a & WN3-4 [MG87]; {\bf WN4b/WCE} [FMG03] & $-$ \\
108 & 89 & 71 & WN7 [Brey]; {\bf WN7h} [CS97] & $-$ \\
136 & 90 & 74 & WC5 [Brey]; {\bf WC4} [SSM90] & $-$ \\
147 & 91 & 73 & WN6.5h and O7~V [W99]; WN6h:a [S08] & {\bf WN6(h)} \\
180 & 93 & \phantom{a}74a & O3~If$^\ast$/WN6 [TS90] & {\bf O3~If$^\ast$} \\
402 & 95 & 80 & WN7 [F60]; WN7 [Brey]; WN6 [Mk]; WN7h [CS97] & {\bf WN7h$+$OB} \\
427 & 96 & 81 & WN8 [Brey]; WN8 [Mk]; {\bf WN8(h)} [CS97] & $-$ \\
457 & 97 & $-$ & O4~If [Mk]; O3 If$^\ast$/WN7-A [WB97] & {\bf O3.5 If$^\ast$/WN7} \\
482 & 99 & 78 & O4~If [Mk]; O3~If$^\ast$/WN6-A [WB97] & {\bf O2.5 If$^\ast$/WN6} \\
507 & 101/102 & 87 &  WC5$+$WN4 [Mk]; {\bf WC4$+$WN6} [M87];  & $-$ \\
509 & 103 & 87 & WN4.5 [Mk]; WN6 [M87]; WN5.5 [B99] & {\bf WN5(h)$+$O} \\
542 & 113 & $-$ & O3 If [Mk]; O3~If$^\ast$/WN6-A [WB97] & {\bf O2~If$^\ast$/WN5} \\
545 & 114 & $-$ & O3 If [Mk]; O3~If$^\ast$/WN6-A [WB97] & {\bf O2 If$^\ast$/WN5} \\
617 & 117 & 88 & {\bf WN5ha} [FMG03] & $-$ \\
682 & $-$ & $-$ & $-$ & {\bf WN5h} \\
695 & 119 & 90 &  WN6-7 [F60]; WN7 [Brey]; WN6 [Mk]; WN6 [B99]; WN6(h) [CS97] & {\bf WN6h$+$?} \\
731 & 121 & \phantom{a}90a & WC4 [MG87]; {\bf WC4} [SSM90]; WC7-9 [B99] & $-$ \\
758 & 122 & 92 & WN5$+$ [F60]; WN6 [Brey]; WN5(h) [FMG03] & {\bf WN5h} \\
\hline
\end{tabular}
\end{center}
\tablefoot{Previous classifications are from F60 \citep{f60}; Brey \citep{brey}; W84 \citep{w84}; 
Mk \citep{m85}; MG87 \citep{mg87}; M87 \citep{m87}; SSM90 \citep{ssm90}; 
TS90 \citep{ts90}; CS97 \citep{cs97}; WB97 \citep{wb97}; B99 \citep{b99}; W99 \citep{wd99}; 
FMG03 \citep{f03}; S08 \citep{schnurr08}.}
\end{table*}

\subsection{Discovery of a new W--R star in the LMC}\label{682}

The combined FLAMES spectrum of VFTS~682 is shown in
Figure~\ref{682_spec}.  Classified as WN5h (in which the `h' suffix
denotes the presence of hydrogen lines), this is a previously unknown
W--R star.  From qualitative comparisons of the individual spectra
there is no evidence for significant ($\Delta v$\,$\gtrsim$\,10\,\kms)
radial velocity variations.

In their analysis of the {\em Spitzer}-SAGE survey of the LMC
\citep{sage}, VFTS~682 is included by \citet[][hereafter GC09]{gc09}
in their list of `definite' young stellar objects (YSOs).  Near-IR
(from the IRSF catalogue) and {\em Spitzer} photometry for VFTS~682
are summarised in Table~\ref{ysomatches}.  Near-IR photometry is also
available from \citet[][infrared source 153]{h92} and 2MASS.  The
evolutionary status of VFTS~682 and the origins of its mid-IR excess 
are discussed by Bestenlehner et al. (in preparation).

\subsubsection{A new B[e]-type star adjacent to R136}

\begin{center}
\begin{figure*}
\includegraphics{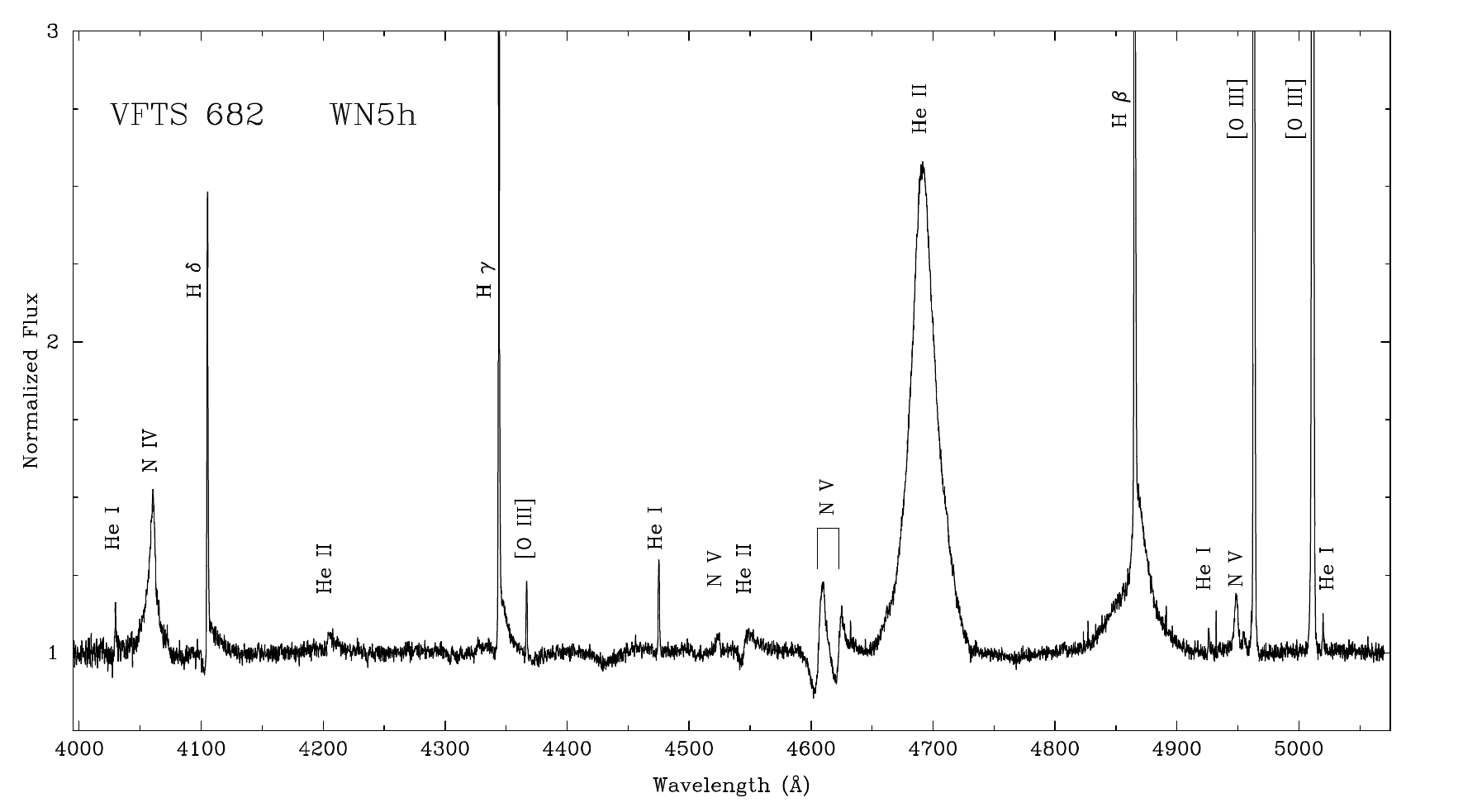}
\caption{Combined FLAMES--Medusa spectrum of VFTS~682, a newly
  discovered Wolf-Rayet star.  Identified lines are, from
  left-to-right by species: H$\delta$, H$\gamma$, and H$\beta$ (each
  with super-imposed nebular emission); He~{\scriptsize{I}}
  \lam\lam4026, 4471, 4922, 5015 (each from nebular emission);
  He~{\scriptsize{II}} \lam\lam4200, 4542, 4686; N~{\scriptsize{IV}}
  \lam4058; N~{\scriptsize{V}} \lam\lam4520, 4604-4620, 4944; and the
  [O~{\scriptsize{III}}] nebular lines at \lam\lam4363, 4959,
  5007.}\label{682_spec}
\includegraphics{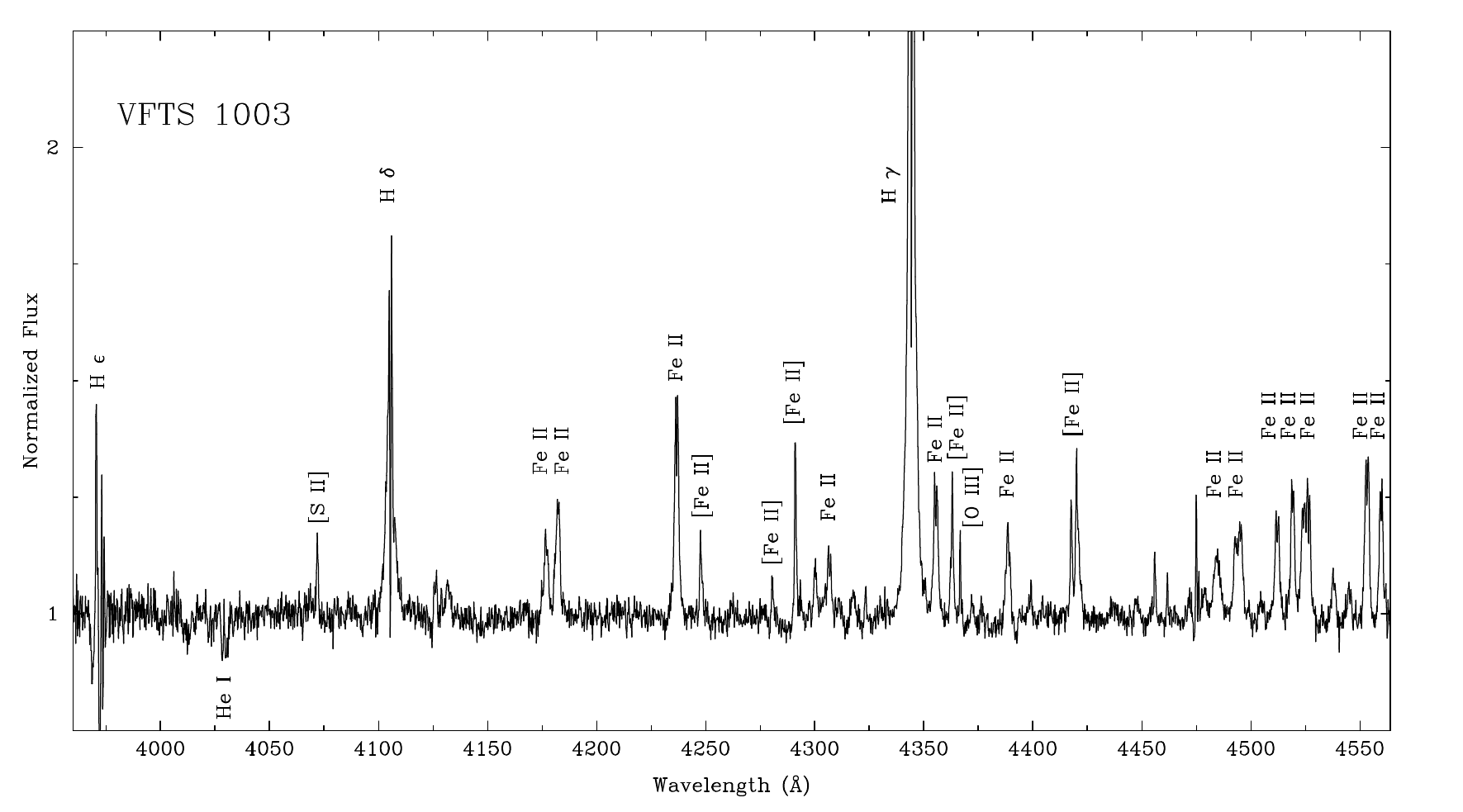}
\caption{Combined FLAMES--ARGUS spectrum of VFTS~1003, a newly
  discovered B[e]-type star.  There appears to be weak stellar
  absorption at He~{\scriptsize{I}} \lam\lam4009, 4026, 4388, 4471,
  combined with weak nebular emission components, particularly at \lam4471.  In
  addition to the Balmer lines, the identified lines are
  [S~{\scriptsize{II}}] \lam4069; [O~{\scriptsize{III}}] \lam4363;
  Fe~{\scriptsize{II}} \lam\lam4173, 4179, 4233, 4303, 4352, 4385, 4481,
  4491, 4508, 4515, 4520-23, 4549, 4556; [Fe~{\scriptsize{II}}]
  \lam\lam4244, 4277, 4287, 4358-59, 4414-16.}\label{1003_spec}
\end{figure*}
\end{center}

The ARGUS spectrum of VFTS~1003 (Figure~\ref{1003_spec}) is
particularly striking owing to a large number of Fe~\2 emission lines.
It is similar to that of GG~Carinae \citep[see][]{wf00} but with emission
lines from [Fe~\2]; some weak He~\1 absorption lines
are also present.  There is forbidden [S~\2] and [O~\3] emission at
\lam4069 and \lam4363, respectively, although from the present data 
it is not clear if these are from the local nebulosity or are intrinsic to the star.
There do not appear to be any significant radial
velocity shifts between the individual spectra.
Its near-IR colours $(J-H)$\,$=$\,0.54 and $(H-K_{\rm
s})$\,$=$\,1.43\,mag (from the IRSF catalogue) place it in a
comparable region to the B[e] stars from \citet{gzw95}.  We also note
that it is a single, isolated source in the high angular-resolution
near-IR images from the Multiconjugate Adaptive-optics Demonstrator
(MAD) from \citet{mad}.

\citet{l98} presented a classification framework for B[e]-type stars, 
with GG~Car classified as a B[e] supergiant. The spectrum of VFTS~1003
warrants a B[e] classification, but does not allow us
to distinguish between an evolved B[e] supergiant and a pre-main
sequence (`Herbig') B[e] star (cf. the criteria from \citeauthor{l98}).
Nevertheless, the presence of such a rare object a mere 8\farcs5 from
the core of R136 certainly warrants further study.  In particular,
photometric monitoring would help to distinguish between the two
evolutionary scenarios: small variations ($\sim$0.2\,mag) would be
expected for a B[e]-type supergiant, whereas much larger and irregular
variations (related to accretion processes) would be expected for a
`Herbig' object \citep{l98}.

\subsection{Classification of later-type stars}\label{coolstars}

There are 91 stars with classifications of A-type or later that were
retained as likely members of the LMC.  For the purposes of
classification these were assumed to be single stars, i.e. all of
the available LR02 and LR03 spectra were stacked and co-added; classifications are
given in Table~\ref{spec_cool}.

The small number of A-type spectra were classified with reference to the
standards presented by \citet{eh03} and the metal-poor A-type
stars from \citet{f2}.  The new FLAMES data do not include the Ca~$K$
line, so the primary temperature diagnostic is the intensity of the
metal lines; luminosity types were assigned on the basis of the
H$\gamma$ equivalent-width criteria from \citet{eh04}.

Except for a small number of G-type stars
\citep[that were classified following the criteria from][]{eh03} we employ
broad classification bins for the cooler types.  These are not a key
component of our scientific motivations and due to their red colours
have relatively low S/N ratios.  The spectral bins adopted in
Table~\ref{spec_cool} encompass a range of types: `Early~G' (G0-G5);
`Late G/Early K' (G5-K3); `Mid-late~K' (K3-M0); and `Early M'.  Three
carbon stars with strong C$_2$ Swan bands are also included
in the sample.

We do not attempt luminosity classifications for stars of F-type and
later.  On the basis that their radial velocities are consistent with
membership of the LMC, from distance arguments they are notionally
supergiants or bright giants (i.e. classes I and II).  Indeed, five
stars in Table~\ref{cat} are sufficiently bright\footnote{Assuming
M$_V$\,$\sim$$-$4.5 to 5.0 for late-type Ib supergiants.} to be
considered `red supergiants': VFTS~081, 198, 236, 289, and 793,
each of which is encircled in red in Figure~\ref{targets}.

\begin{table*}
  \caption{Spectral classifications for cool-type stars (A-type or later) 
from the VLT-FLAMES Tarantula Survey (VFTS).}\label{spec_cool}
\begin{center}
\begin{tabular}{llcllcll}
\hline\hline
Star & Sp. Type & ~~~~ & Star & Sp. Type & ~~~~ & Star & Sp. Type \\
\hline
006 & Mid-late K        &&323 & A5 II             &&773 & Late G/Early K    \\
011 & Late G/Early K    &&341 & Mid-late K        &&776 & Carbon star       \\
023 & Late G/Early K    &&344 & Mid-late K        &&783 & Late G/Early K    \\
026 & Late G/Early K    &&357 & Late G/Early K    &&785 & Mid-late K        \\
032 & Late G/Early K    &&372 & Late G/Early K    &&790 & F0                \\
057 & Mid-late K        &&379 & Mid-late K        &&791 & Late G/Early K    \\
081 & Mid-late K        &&437 & Mid-late K        &&793 & Late G/Early K    \\
092 & Late F            &&439 & Late G/Early K    &&803 & Late G/Early K    \\
115 & Late G/Early K    &&454 & Late G/Early K    &&805 & Mid-late K        \\
129 & Mid-late K        &&490 & Late G/Early K    &&808 & Late G/Early K    \\
139 & Late F            &&524 & G0                &&809 & Late G/Early K    \\
175 & A2-3 II           &&544 & Late G/Early K    &&816 & G2                \\
182 & F0                &&595 & Mid-late K        &&818 & Mid-late K        \\
193 & Late G/Early K    &&614 & Early G           &&820 & A0 Ia             \\
198 & Mid-late K        &&655 & Late G/Early K    &&828 & Early M           \\
222 & G0                &&658 & A7 II             &&839 & G                 \\
236 & Mid-late K        &&674 & A2-3 II           &&844 & Mid-late K        \\
245 & Mid-late K        &&680 & Early G           &&852 & Late F            \\
260 & Early G           &&691 & A2-3 II           &&856 & A7 II             \\
262 & F0                &&693 & Late G/Early K    &&858 & A7 II             \\
264 & Mid-late K        &&694 & Mid-late K        &&861 & Late G/Early K    \\
265 & A2-3 II           &&700 & Mid-late K        &&862 & Early G           \\
271 & A7 II             &&708 & Late G/Early K    &&863 & A5 II             \\
275 & Early M           &&721 & A9 II             &&865 & Late G/Early K    \\
281 & Mid-late K        &&744 & Early M           &&870 & F0                \\
289 & Late G/Early K    &&759 & Mid-late K        &&871 & A7 II             \\
294 & A0 Ib             &&760 & A9-F0 II          &&873 & A2-3 II           \\
311 & Carbon star       &&763 & G5:               &&878 & G2                \\
312 & Mid-late K        &&765 & Early M           &&884 & Carbon star       \\
317 & A9 II             &&767 & Late G/Early K    &&893 & A7: II            \\
319 & Mid-late K        &&&&&&\\
\hline
\end{tabular}
\end{center}
\tablefoot{Previous classifications are available for 11 stars, primarily
  from \citet[][Mk]{m85} and \citet[][ST92]{st92}: 057: G8-K2 [ST92]; 
  129: G8-K3 [ST92]; 193: late G [ST92]; 198: G,K [ST92]; 222: G,K [ST92]; 
  271: A~Ib [Mk]; 294: A0~Ib [Mk]; 317: A5~Ib
  [Mk]; 437: G8~III \citep{b99}; 691: A~I \citep{wb97}, A2-3~I \citep{b99}; 
  793: early K I \citep{p93}.}
\end{table*}

\section{{\em Spitzer} YSO candidates}\label{yso}

Prompted by the mid-IR behaviour of VFTS~682 (Section~\ref{682}), we
cross-matched the survey targets with the catalogues of `definite' and
`probable' YSOs from GC09 to investigate their spectral properties.

The SAGE data comprise imaging with two {\em Spitzer}
instruments: the InfraRed Array Camera (IRAC) at 3.6, 4.5, 5.8, and
8.0\,$\mu$m, and the Multiband Imaging Photometer for Spitzer (MIPS)
at 24, 70, and 160\,$\mu$m.  The angular resolution of the IRAC images
ranges from 1\farcs7 to 2\farcs0, with coarser resolution of 6$''$,
18$''$, and 40$''$, at 24, 70, and 160\,$\mu$m, respectively
\citep{sage}.  To compare our targets with the GC09 catalogues, we employed a
conservative search radius (compared to the IRAC resolution) of
$r$\,$<$\,2\farcs5, which yielded five potential matches in the `definite'
YSO list, and 11 potential matches in the `probable' YSO list.  Each
potential match was then examined using the optical WFI images and,
where possible, the MAD near-IR imaging from
\citet{mad}, and the imaging from the High Acuity Wide-field K-band
Imager (HAWK-I) which was used to calibrate the MAD data.

The near-IR images are particularly helpful to identify robust
counterparts.  A notable false match at a distance of 1\farcs2 from
VFTS~476 is GC09 {053839.69$-$690538.1}.  The optical WFI image
reveals only the VFTS target whereas, as shown by Figure~13 from
\citet{mad}, there is a very red source at the GC09 position, with
VFTS~476 the object to the south\footnote{\citet{mad} argued that the
  star was likely a massive star on the basis of its near-IR photometry.
  Indeed, the FLAMES spectra of VFTS~476 reveal it as a late O-type
  star.}.  Near- and mid-IR photometry of matched sources from GC09
are summarised in Table~\ref{ysomatches}, each of these is now
discussed in turn.\\

{\bf VFTS~016:} This is the massive `runaway' star from
\citet[][`30~Dor~016', classified as O2~III-If$^\ast$]{e10}, which
features in the list of probable YSOs from GC09.

Mid-IR observations have been used by, e.g., \citet{gb08} and \citet{g10,g11} 
to identify bow shocks associated with runaway stars.  For instance, the SAGE
24\,$\mu$m images were used to investigate six O2-type stars
in the LMC thought to be runaways, including VFTS~016 \citep{g10}.
Of the six candidate runaways, only BI\,237 was reported to have a bow shock from
inspection of the MIPS images, consistent with the expectation from
\citet{gb08} that approximately 20\% of runaways have an associated
bow shock.

Although the {\em Spitzer} resolution at 24\,$\mu$m is less than ideal
in a region as crowded as 30~Dor, VFTS~016 is relatively isolated at a
projected distance of 120\,pc from the core of R136.  The
24\,$\mu$m magnitude from GC09 for VFTS~016 (see
Table~\ref{ysomatches}) suggests a strong mid-IR excess -- perhaps
associated with a bow shock (but not extended sufficiently to be
detected by \citeauthor{g10}).

\begin{center}
\begin{figure*}
\includegraphics{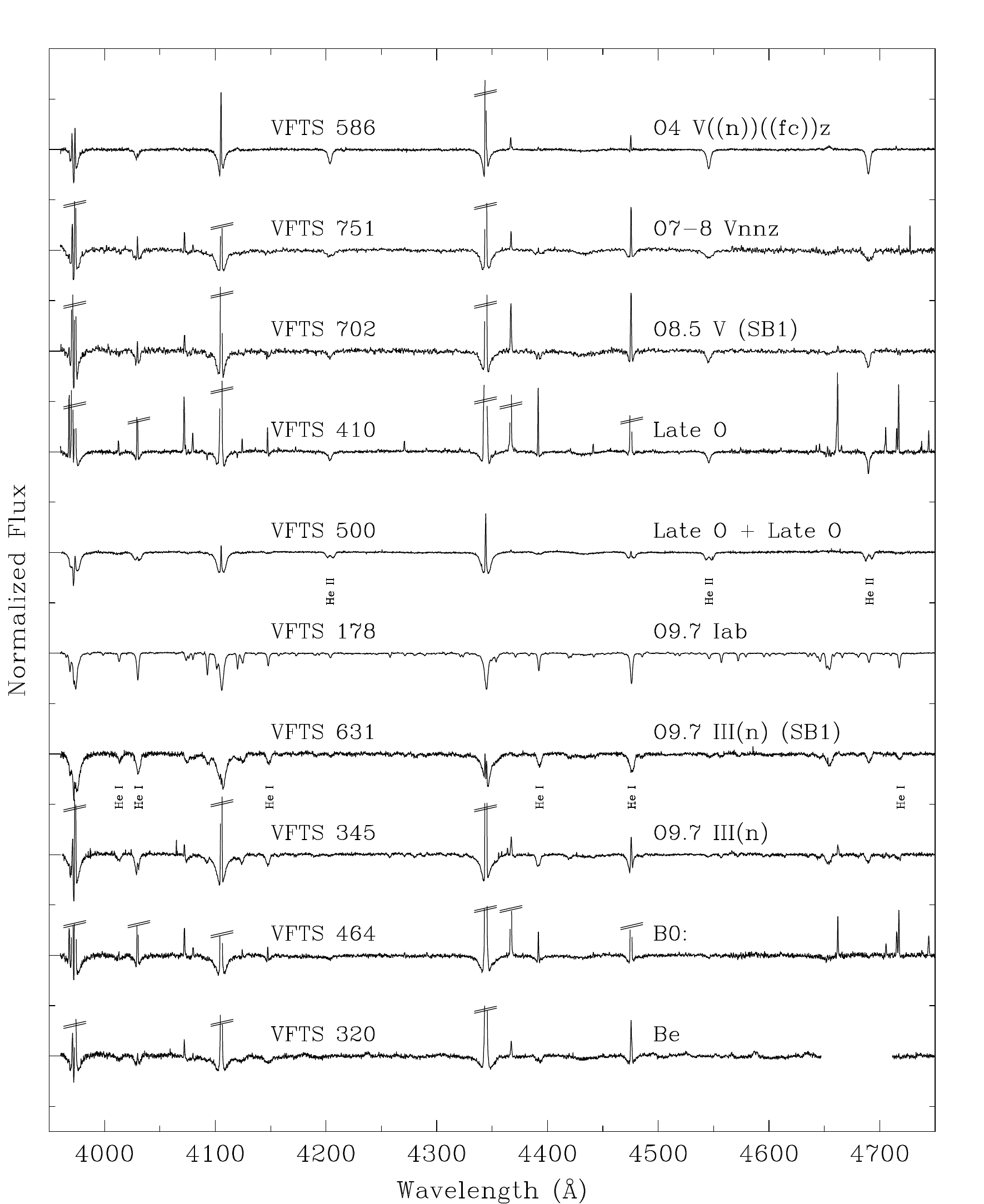}
\caption{FLAMES--Giraffe spectroscopy of ten candidate YSOs from \citet{gc09}, as summarised in 
Table~\ref{ysomatches}.  Strong nebular lines have been truncated as
indicated. The He~{\scriptsize{II}} lines identifed in VFTS~500 
(a double-lined spectroscopic binary) are: \lam\lam4200, 4542, 4686;
He~{\scriptsize{I}} lines identified in VFTS~631 are: \lam\lam4009, 4026, 4144, 4388, 4471, 4713.}\label{montage}
\end{figure*}
\end{center}

{\bf VFTS~178:} A visually bright star, the spectrum of VFTS~178 appears as a
relatively unremarkable supergiant, classified O9.7~Iab (see
Figure~\ref{montage}).  The star was previously classified as
B0.5~I by \citet{st92}, with the new high-quality data revealing
sufficiently strong He~\2 \lam4542 absorption that a slightly earlier
type is required.  No obvious radial velocity variations are apparent
from inspection of the FLAMES data.\\

{\bf VFTS~320:}  This was the closest match to GC09 {053821.10$-$690617.2}
(with VFTS~316 at a distance of only 1\farcs2).  As noted in
Table~\ref{cat}, the spectrum of VFTS~320 suffers from contamination
from an adjacent fibre on the detector (see Section~\ref{redux}),
manifested by a broad emission bump (approximately 4\% above the
continuum) at \lam4686.  Aside from this wavelength region, the spectrum 
is that of an early B-type star with considerable nebular
contamination plus weak Fe~\2 emission, which resembles that from a 
Be-type star.  The adjacent object on the detector was VFTS~147, classified in
Section~\ref{wrstars} as WN6(h).  This does not display Fe~\2 emission
(indeed, the He~\2 \lam4686 emission is by far the strongest line),
i.e., the Fe~\2 features seen in VFTS~320 are genuine.  The spectrum (with
the \lam4686 region omitted) is shown in Figure~\ref{montage}.\\

{\bf VFTS~345:} A single object in the WFI and HAWK-I frames, the
co-added spectrum of VFTS~345 is shown in Figure~\ref{montage} and
is classified as O9.7~III(n) \citep[cf. the published classification of B0~V from][]{b99}.\\

{\bf VFTS~410:} This target is P93-409 \citep{p93}, located in `Knot~3' to 
the west of R136 \citep{w91}.  Classified as O3-6~V by \citet{wb97}, {\em
HST} imaging and spectroscopy by \citet{w99,wmb02} resolved two bright
components with spectral types of O8.5~V and O9~V. A third, much
redder (visually fainter) source was reported just 0\farcs4 northwest
of the two massive stars by \citet{w99}, as well as several very
compact nebulosities to the northeast.  This complex would be resolved
poorly by the {\em Spitzer} observations, but is clearly an active
site of star formation.

For completeness, we show the combined spectrum of VFTS~410 in
Figure~\ref{montage}; there is significant nebular emission
superimposed on the stellar profiles.  Given the knowledge that there
are two luminous components contributing to these data, it is worth
noting that no radial velocity variations are seen from a comparison of
the available spectra.  Even if these are not in a bound binary
system, their proximity in such a star-formation complex suggests they
are associated.  This object highlights the benefit of high-resolution
imaging for some systems (from {\em HST}, MAD, etc.) when attempting
to interpret the new spectroscopy.\\

\begin{figure}
\begin{center}
\includegraphics[height=7.35cm]{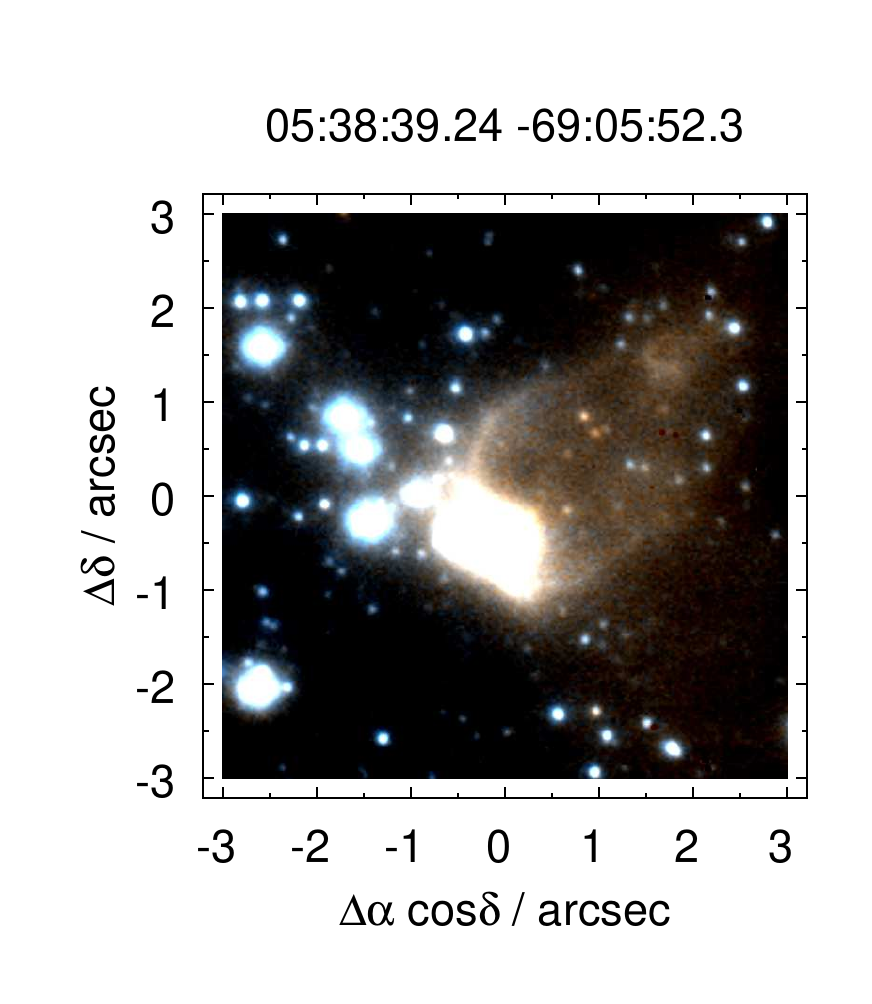}\\
\includegraphics[height=7.35cm]{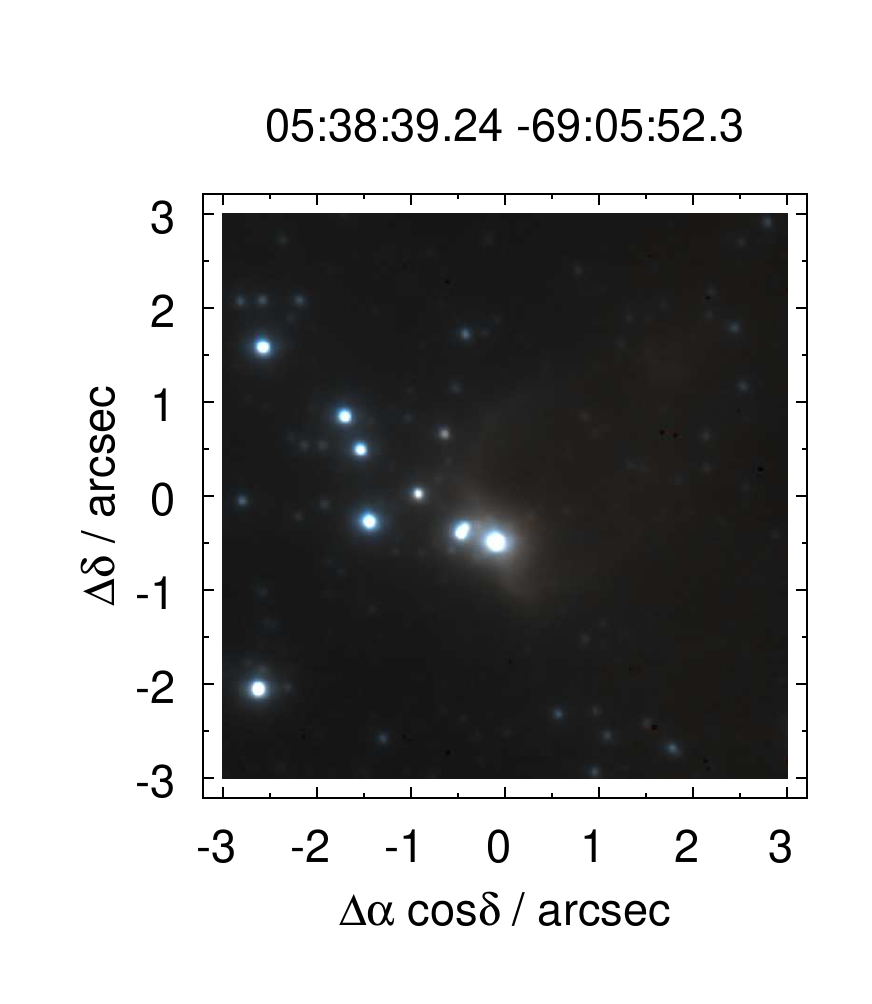}
\caption{Combined 6$''$\,$\times$\,6$''$ $H$- and $K_{\rm s}$-band MAD image
  of the candidate young stellar object 053839.24$-$690552.3 from the catalogue of \citet[][centred on their
  position]{gc09}.  The intensity is scaled to bring out the bow shock and the source at its centre, 
  VFTS~464, in the upper and lower panels, respectively.}\label{yso_464}
\end{center}
\end{figure}

{\bf VFTS~464:} This is the star at the centre of the impressive
bow-shock feature discovered by \citet{mad} from the MAD images.  In
Figure~\ref{yso_464} we show new combined colour images from MAD
(with different intensity scalings).  There is a source 0\farcs4 from
the central object that is approximately three times fainter -- given
the 1\farcs2 aperture of the Medusa fibres, the spectra of the central
star are likely contaminated by the companion.

As one might expect, the FLAMES spectra are heavily contaminated by
nebular emission, which appears to include forbidden lines such as
[S~\2] \lam\lam4069-76, [Fe~\3] \lam4702, and [Ar~\4] \lam4711.  The
combined blue-region spectrum is shown in Figure~\ref{montage} -- note
the weak He~\2 absorption, leading to a provisional classification of
B0:, with the uncertainty reflecting the problems of nebular
contamination and the potential contribution from the nearby object.  

The He~\2 absorption at \lam4542 and \lam4686 appears consistent with
the systemic radial velocity of 30~Dor, i.e., the object does
not appear to be a candidate runaway in the radial direction.  Indeed,
as noted by \citet{mad}, the bow shock is orientated {\em towards}
R136 (and Brey~75/BAT99-100), suggesting it might well be related to an
ionization front (with associated triggered star-formation) rather than a 
dynamical shock.\\

\begin{figure}
\begin{center}
\includegraphics[height=7.35cm]{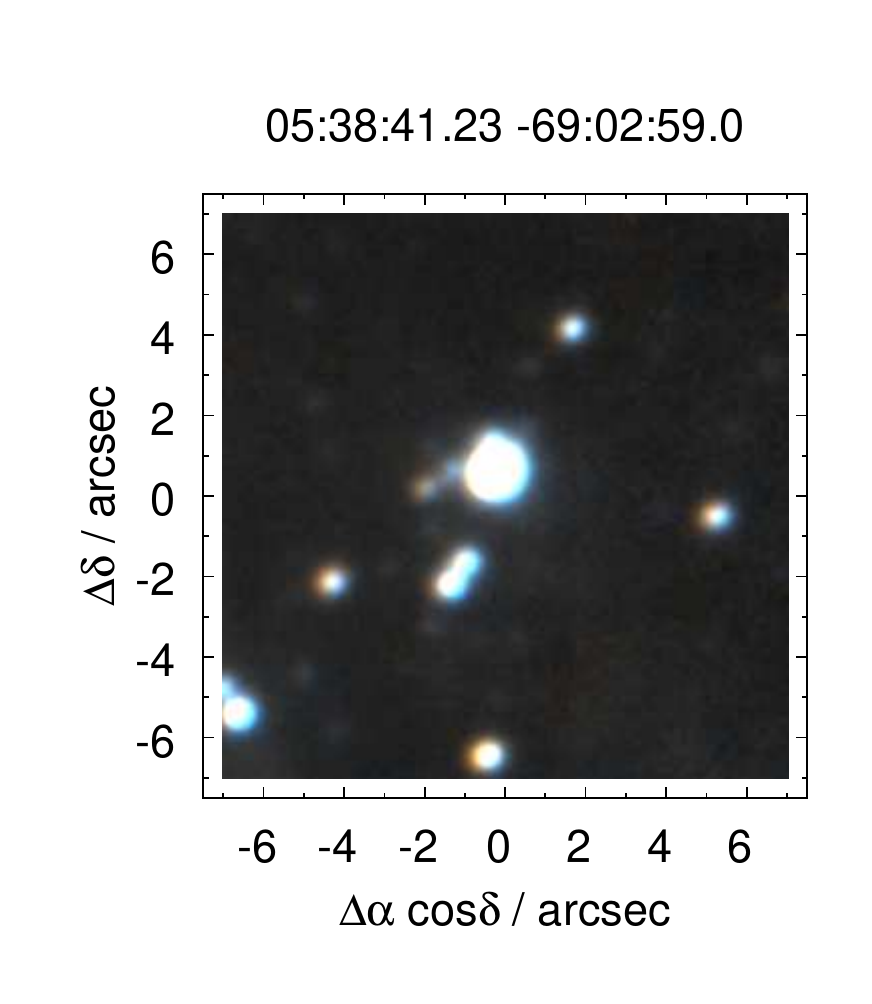}
\caption{Combined 14$''$\,$\times$\,14$''$
$J$- and $K_{\rm s}$-band HAWK-I image of the candidate young stellar object 
053841.23$-$690259.0, centred at the position from \citet{gc09}.
The (near-) central bright source is VFTS~500, a double-lined binary comprised
of two late O-type stars.}\label{yso_500}
\includegraphics[height=7.35cm]{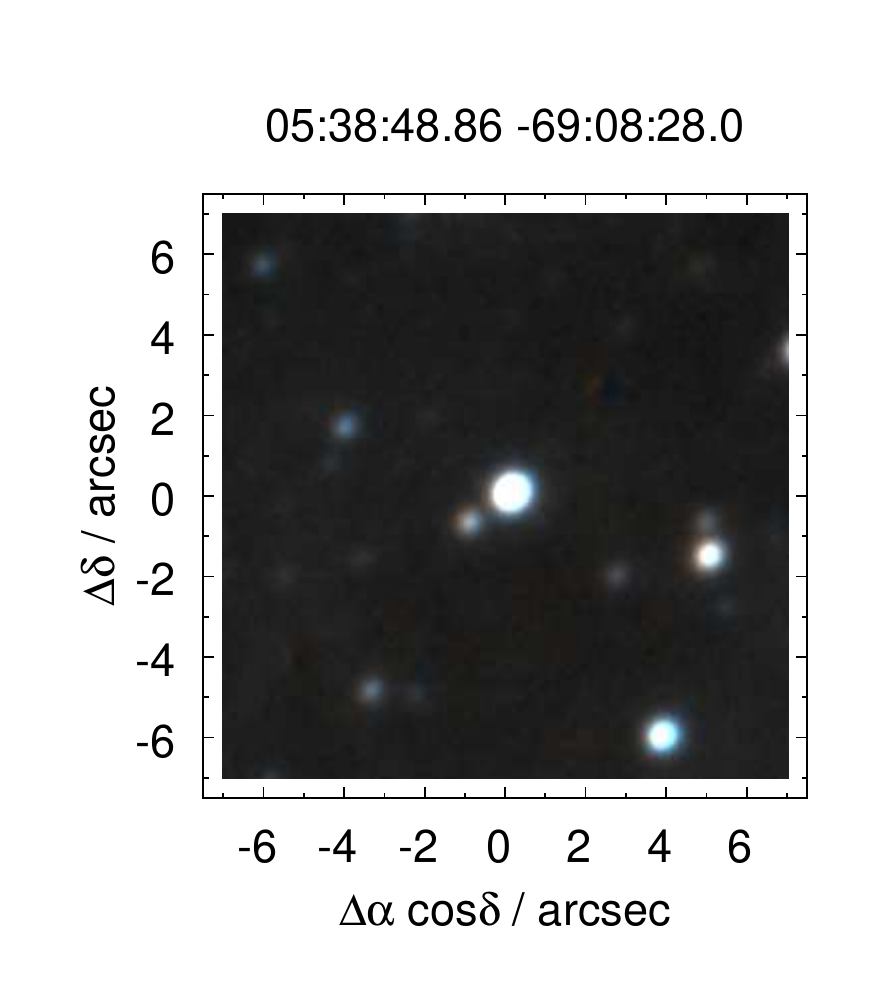}
  \caption{Combined 14$''$\,$\times$\,14$''$ $J$- and $K_{\rm s}$-band
  HAWK-I image of the candidate
  young stellar object 053848.86$-$690828.0, centred at the position from \citet{gc09}.  The
  (near-) central bright source is VFTS~631, a single-lined massive
  binary.}\label{yso_631}
\end{center}
\end{figure}

{\bf VFTS~500:} This object appears as a single source in the HAWK-I imaging, as shown
in Figure~\ref{yso_500} (in which the slight northern extension of the
star is an artefact of the dither pattern used for the
observations)\footnote{The two sources $\sim$2\farcs75 to the SSE
comprise VFTS~504, and the two at the SE edge are VFTS~530.  Both
appear as probable blends in the WFI imaging.}; the FLAMES
spectroscopy reveals it as a double-lined binary.  Observed in Medusa
Field~A, the first LR03 epoch was obtained on the same night as the
first three LR02 spectra.  The co-added spectrum from these epochs is
shown in Figure~\ref{montage}, displaying twin components in the He~\1
and \2 lines.  The nebular contamination thwarts precise
classification, but both of the He~\1 \lam4026 components are greater
in intensity than the He~\2 \lam4200 features, requiring a type of
later than O6.  Similarly, the intensity of the He~\2 \lam4200
profiles suggests a type of earlier than B0 for both components.\\

{\bf VFTS~586:} The combined blue-region spectrum is shown in Figure~\ref{montage}.
The spectrum is classified O4~V((n)))((fc))z, in which the `c' suffix
refers to C~\3 emission at \lam\lam4647-50-52 \citep{w10}.\\

{\bf VFTS~631:} Inspection of the spectra of VFTS~631 reveals a single-lined binary,
with radial velocity shifts of the order of 70\,\kms in the final LR02
observation compared to the first epoch.  To estimate the spectral
type, the first three LR02 epochs (taken on the same night) were
combined with the LR03 data, as shown in Figure~\ref{montage}. The
combined spectrum is classified as O9.7~III(n), in good agreement with published
types of O9-B0~II \citep{wb97} and O9.5~II \citep{b99}.

There are two obvious components to the nebular emission in the LR02
and LR03 spectra (e.g. twin-peaked emission in the [O~\3] lines).  One
of the components is consistent with the typical systemic velocity of
30~Dor ($\sim$270--280\,\kms, with the other blueshifted by
approximately 50--60\,\kms).  These are most likely from separate
components of gas emission, but could also be indicative of a
wind-blown bubble around the binary, depending on the systemic velocity
of the system. A third (weaker), longer-wavelength nebular component is also visible in
the [N~\2] and [S~\2] lines in the HR15N spectra.\\

{\bf VFTS~702:} As with VFTS~631, radial velocity shifts are seen between the LR02 
observations. To estimate the spectral type we combined
the second and third LR02 epochs with the LR03 observations (all
obtained on the same night), shown in Figure~\ref{montage} and 
classified as O8.5~V.\\

{\bf VFTS~751:} An apparently single star, with the combined spectrum shown in
Figure~\ref{montage}.  The stellar lines appear broadened by rotation, leading to a 
classification of O7-8~Vnnz.

\begin{table*}
  \caption{Summary of matches between targets in the VLT-FLAMES
  Tarantula Survey (VFTS) and candidate young stellar objects (YSOs)
  from \citet[][`GC09']{gc09}.  The fourth column is the source
  classification from GC09: `C` -- candidate YSO; `CS' -- probable
  YSO, but could possibly be a star; `SD' -- stars with photometric
  contamination from diffuse emission. Near-IR photometry ($JHK_{\rm
  s}$) is from the IRSF catalogue \citep{irsf}, with mid-IR
  photometry from GC09.}\label{ysomatches}
\begin{center}
\begin{tabular}{clccccccccccc}
\hline\hline
Star & VFTS Sp. Type & Identification & Type & $\Delta r$ & $J$ & $H$ & $K_s$ & [3.6] & [4.5] & [5.8] & [8.0] & [24.0] \\
{[VFTS]} & & [GC09] & & [$''$] & & & & & & & & \\
\hline
016 & O2 III-If$^\ast$ & 053708.79$-$690720.3 & CS & 0.5 & 13.39 & 13.37 & 13.35 & 12.99 & 12.20 & 10.61 & \oo8.73 & 3.82 \\
178 & O9.7 Iab & 053751.03$-$690934.0 & CS & 0.1 & 12.84 & 12.83 & 12.81 & 12.71 & 12.48 & 11.69 & \oo9.76 & 3.87 \\
320 & Be & 053821.10$-$690617.2 & CS & 0.4 & 15.19 & 14.98 & 14.64 & 13.21 & 12.86 & 11.85 & 10.04 & $-$ \\
345 & O9.7 III(n) & 053827.39$-$690809.0 & CS & 0.4 & 15.48 & 15.47 & 15.37 & 13.82 & 12.88 & 11.48 & \oo9.70 & $-$ \\
410 & Late O & 053834.77$-$690606.1& C & 0.9 & \multicolumn{8}{c}{Multiple components -- see \citet{w99,wmb02}} \\
464 & B0: & 053839.24$-$690552.3 & C & 1.2 & 14.55 & 14.26 & 13.63 & 11.27 & 10.40 & \oo8.98 & \oo6.93 & $-$ \\
500 & Late O$+$ late O & 053841.23$-$690259.0 & CS & 0.7 & 13.80 & 13.73 & 13.68 & 13.27 & 12.52 & 10.89 &  \oo8.74 & 3.02 \\
586 & O4 V((n))((fc))z & 053845.33$-$690251.6 & CS & 0.4 & 14.96 & 14.91 & 14.88 & 14.51 & 13.34 & 11.62 & \oo9.51 & 2.82 \\
631 & O9.7 III(n) (SB1) & 053848.86$-$690828.0 & CS & 0.1 & 15.76 & 15.66 & 15.66 & 14.60 & 13.53 & 12.37 & 11.13 & $-$ \\
682 & WN5h & 053855.56$-$690426.5 & CS & 0.3 & 13.53 & 12.96 & 12.48 & 11.69 & 11.37 & 10.48 & \oo9.15 & $-$ \\
702 & O8.5 V (SB1) & 053858.42$-$690434.7 & C & 0.5  & 15.35 & 15.12 & 14.91 & 12.57 & 11.74 & \oo9.59 & \oo7.71 & $-$ \\
751 & O7-8 Vnnz & 053909.13$-$690128.7 & CS & 0.9 & 15.62 & 15.48 & 15.31 & 13.87 & 13.07 & $-$ & 10.37 & 3.75 \\
\hline
152 & B2 V & 053746.64$-$690619.8 & SD & 0.4 & 14.66 & 14.47 & 14.30 & 14.05 & 13.99 & 13.26 & 11.76 & $-$ \\
323 & A5 II & 053821.78$-$691042.9 & SD & 0.1 & 14.26 & 14.07 & 13.91 & 13.81 & 13.86 & 13.15 & 11.57 & $-$ \\
641 & B0.5: I: & 053849.81$-$690643.3 & SD & 0.5 & 12.63 & 12.58 & 12.48 & 11.46 & 10.54 & \oo9.67 & \oo7.34 & 0.06 \\
842 & Be & 053942.66$-$691151.6 & SD & 0.3 & 15.08 & 14.85 & 14.53 & 13.58 & 13.40 & $-$ & 10.98 & $-$ \\
\hline
\end{tabular}
\end{center}
\tablefoot{
VFTS~410\,$=$\,P93-409; the two major components were classified by \citet{wmb02} as O8.5~V and O9~V.}
\end{table*}

\subsection{Discussion}

All of the spectra in Figure~\ref{montage} (except VFTS~178) contain
some degree of nebular contribution; this is not surprising given the spatial
extent of the H~\2 region in 30~Dor.  However, the most intense [O~\3]
emission is seen in VFTS~410, 464, and 702, the three sources
indentified by GC09 as clear YSO candidates.  Both VFTS~410 and 464
are located immediately to the west of R136, while VFTS~702 is 2$'$ to
the northeast.  Echoing the discussion from
\citet{mad} regarding the location of the candidate high-mass YSOs
(which included VFTS~464), all three objects are in regions associated
with molecular gas \citep{w78,j98} that comprise the second
generation of star-formation around R136
\citep{wb97,w99,wmb02}.  

Closer scrutiny of the correlation between H$\alpha$ emission (from
archival {\em HST} images) and the GC09 sources was provided by
\citet{v09}, in which they discussed four of the VFTS sources: VFTS~464 
was identified as a Type~II YSO \citep[cf. the criteria from][]{chen09} in a bright-rimmed dust
pillar, and VFTS~320, 682 and 702 were each classifed as Type III YSOs
in H~\2 regions.  \citeauthor{v09} noted each of the latter three as
comprising multiple sources in the {\em Spitzer} PSF.  The close
companions to VFTS~320 have already been noted above, while there are
no obvious nearby companions to VFTS~702 in the HAWK-I images, nor in
the IRSF catalogue.  There is a faint IRSF source approximately 2$''$
from VFTS~682: 05385578$-$6904285, with
$H$\,$=$\,17.56\,$\pm$\,0.09\,mag and no detections in $J$ and $K_{\rm
s}$.  Intriguingly, VFTS~682 is only $\sim$17\farcs5 from VFTS~702,
both of which appear to be co-located with the most intense
CO-emission in the region \citep[see Fig.~1 from][]{j98}.  This poses
additional evolutionary questions regarding VFTS~682 (Bestenlehner et
al. in preparation).

From their multi-band approach, \citet{v09} reported VFTS~016 and 631
as non-YSOs.  They describe VFTS~016 (GC09\,053708.79$-$690720.3) as a
bright star, consistent with the mid-IR excess perhaps arising from a
bow shock.  However, they flag VFTS~631 (GC09\,053848.86$-$690828.0)
as a galaxy.  Notwithstanding the discussion of the spectroscopy of
VFTS~631, the combined $J$- and $K_{\rm s}$-band HAWK-I image appears
unremarkable (Figure~\ref{yso_631}), nor is it obviously extended in
the WFI images.

Analysis of the SAGE data by \cite{whitney} identified 1197 candidate
YSOs using different selection criteria to GC09.  The relative merits
of the selection criteria were discussed by GC09, who note that 72.5\%
of their `probable' YSOs were not in the \citeauthor{whitney}
catalogue owing to conservative identification criteria.  While the nine
probable YSOs from GC09 with VFTS spectroscopy in
Table~\ref{ysomatches} are all interesting objects in the context of
their mid-IR behaviour, none appear to be genuine high-mass YSOs.
However, two of the more secure candidates (VFTS~410 and 464) appear
to be genuine high-mass YSOs, and are not included in the \citeauthor{whitney}
catalogue.  Indeed, there is only one matched source between the VFTS 
targets and those from \citeauthor{whitney}: VFTS~178.

One potential explanation for the photometric properties of some of
the (non-YSO) stars in Table~\ref{ysomatches} is that they are LMC
analogues of the `dusty' B-type stars with 24\,$\mu$m excesses
discovered by \citet{b07} in the SMC. They suggested that these might
be caused by remnant accretion discs, planetary debris discs, or hot
spots in the local interstellar cirrus; similar excesses for SMC stars
have also been found by \citet{i10} and \cite{ab_smc}.  However, \citet{b07} report
$K_{\rm s}$\,$-$\,[24] colours of $\sim$6.5 mag.  In contrast, the sources
with 24\,$\mu$m detections in Table~\ref{ysomatches} have larger
$K_{\rm s}$\,$-$\,[24] colours.  

A range of physical explanations likely lie behind the mid-IR excesses
in the sources listed in Table~\ref{ysomatches}.  One plausible
explanation for some of the otherwise normal O- and early B-type stars
could be the dusty shocks seen to be associated with massive stars in
the Carina nebula \citep{ns10}. At the distance of the LMC these
diffuse emission regions, depending on their spatial scales, can
sometimes be within the resolution of the {\em Spitzer} observations.

Also included in Table~\ref{ysomatches} are four VFTS targets that
GC09 classified as stellar sources with mid-IR excesses, VFTS~152
(classified here as B2~V), VFTS~323 (A5~II), VFTS~641 (B0.5: I:), and
VFTS~842 (Be).  VFTS~641 \citep[{\em aka} Mk\,11][]{m85} is
particularly bright at 24\,$\mu$m ([24]\,$=$\,0.06\,$\pm$\,0.12\,mag
from GC09), but we note that VFTS~655 is nearby ($\sim$8$''$, cf. the
angular resolution at 24\,$\mu$m of 6$''$) and is classified as late
G/early K (Table ~\ref{spec_cool}), so confusion/crowding might be a
factor.  We only have UVES spectroscopy of VFTS~641, so the
uncertainty in its classification arises from the lack of observations
around 4100\,\AA; nevertheless this agrees well with previous
classifications: B0.5~Ia \citep{m85}; B0-0.5~Ia \citep{w86}; B0~Ib
\citep{wb97}; B0.2~IIII \citep{b99}.

\section{Summary}\label{summary}

We have introduced the VLT-FLAMES Tarantula Survey, which has provided
high-quality spectroscopy of over 800 massive stars in the 30~Doradus region
of the LMC. The survey targets are presented in Table~\ref{cat},
including $B$- and $V$-band optical photometry, and extensive
cross-references with previous identifications in this well-studied
region.  Near-IR ($JHK_{\rm s}$) magnitudes for our targets from the
IRSF catalogue by \citet{irsf} are given in Table~\ref{irsfcat}.

Spectral classifications are given for the massive emission-line stars
in Section~\ref{wrstars}, including the discovery of a new WN5h star
(VFTS~682) 2$'$ from R136, and a new B[e]-type star (VFTS~1003) just
8\farcs5 to the west of R136.  Classifications are also given for the
cool-type stars observed by the survey that have radial velocities
consistent with them being members of the LMC.

In Section~\ref{yso} we investigated the spectral
properties of 12 stars identified as definite or probable YSOs
by GC09, finding an eclectic mixture of objects. VFTS~410 and 464
appear to be embedded massive stars, (i.e. {\em bona fide} high-mass
YSOs), but others include the runaway star (VFTS~016), the new W--R
star (VFTS~682), and three massive binaries.  There are four additional
stars in the survey with mid-IR exceses flagged by GC09, one of
which is VFTS~641 (Mk\,11), an early B-type supergiant that is 
particularly bright at 24\,$\mu$m.

Analysis of the stellar radial velocities and multiplicity of the
high-mass stars in the survey will be presented by Sana et al. and
Dunstall et al. (both in preparation), but the power of the
multi-epoch strategy has already been illustrated by two recent
results.  Namely, the discovery of VFTS~527 (R139) as an intriguing
OIf$+$OIfc binary system \citep{t11}, and the lack of a detected companion
to VFTS~016, which suggest that its peculiar radial velocity 
is caused by it being a massive runaway from the core of 30~Dor \citep{e10}.

Analysis of the spectroscopy will be complemented in the future by two
additional sources of photometry.  Monitoring of selected fields
across 30~Dor is part of an ongoing variability campaign with the 2\,m
Faulkes Telescope South at the Siding Spring Observatory, Australia.
This builds on previous (and still ongoing) follow-up of the FLAMES
fields from \citet{f2} to determine periods for notable binaries
(e.g. Ritchie et al. in preparation).  The 30~Dor region was also one of the first
fields observed by the ongoing VISTA Magellanic Clouds Survey
\citep{vmc}, with time-linked $K_{\rm s}$-band observations
to investigate variability.\\

The Tarantula Survey is by far the largest homogeneous spectroscopic
study of extragalactic early-type stars undertaken to date.  The
30~Dor region is the only `super star-cluster' at a well-known
distance in which individual objects can be resolved spatially in
optical light.  This makes it the perfect target for the comprehensive
studies required to address some of the fundamental questions that
remain in our understanding of massive-star evolution, relying on the
analysis of a statistically-significant and unbiased sample.  The
results of these studies will be presented in a series of forthcoming
papers.

\acknowledgements{Based on observations at the European Southern
  Observatory Very Large Telescope in programme 182.D-0222.  We are
  indebted to Claudio Melo and the other ESO staff who have provided
  invaluable assistance with this programme, and to Brian Skiff for
  his dedicated and careful refinements of the published source
  catalogues. We are also grateful to Yuri Beletsky for his
  astrometric calibration, to Simone Zaggia for his advice on the WFI
  data, to Ian Hunter for his input to the original proposal, and to the referee 
  for their helpful comments.  STScI is operated by the Association of Universities for Research in
  Astronomy, under NASA contract NAS\,5-26555, and SdM acknowledges
  NASA Hubble Fellowship grant HST-HF-51270.01-A awarded by STScI.  MG
  acknowledges the Royal Society for financial support.  AH and SSD
  acknowledge funding from the Spanish MICINN (AYA2008-06166-03-01,
  AYA2010-21697-C05-04, Consolider Ingenio 2010 CSD2006-70) and
  Gobierno de Canarias (ProID2010119).  JSV and PRD acknowledge
  support from the Science and Technology Facilities Council.  NM
  acknowledges the Bulgarian National Science Fund under grant
  DO02-85.  VHB acknowledges support from the Scottish Universities
  Physics Alliance (SUPA) and the Natural Sciences and Engineering
  Research Council of Canada (NSERC).  JMA acknowledges support from
  the Spanish Government through grants AYA2010-15081 and
  AYA2010-17631 and the Junta de Andaluc\'{\i}a through grant
  P08-TIC-4075.  AZB acknowledges research and travel support from the
  European Commission FP7 under Marie Curie International
  Re-integration Grant PIRG04-GA-2008-239335.

\bibliographystyle{aa}
\bibliography{16782}

\begin{thebibliography}{110}
\expandafter\ifx\csname natexlab\endcsname\relax\def\natexlab#1{#1}\fi

\bibitem[{{Azzopardi} \& {Breysacher}(1979)}]{ab79}
{Azzopardi}, M. \& {Breysacher}, J. 1979, A\&A, 75, 243

\bibitem[{{Azzopardi} \& {Breysacher}(1980)}]{ab80}
{Azzopardi}, M. \& {Breysacher}, J. 1980, A\&AS, 39, 19

\bibitem[{{Bastian} {et~al.}(2006){Bastian}, {Emsellem}, {Kissler-Patig}, \&
  {Maraston}}]{b06}
{Bastian}, N., {Emsellem}, E., {Kissler-Patig}, M., \& {Maraston}, C. 2006,
  A\&A, 445, 471

\bibitem[{{Bolatto} {et~al.}(2007){Bolatto}, {Simon}, {Stanimirovi{\'c}}, {van
  Loon}, {Shah}, {Venn}, {Leroy}, {Sandstrom}, {Jackson}, {Israel}, {Li},
  {Staveley-Smith}, {Bot}, {Boulanger}, \& {Rubio}}]{b07}
{Bolatto}, A.~D., {Simon}, J.~D., {Stanimirovi{\'c}}, S., {et~al.} 2007, ApJ,
  655, 212

\bibitem[{{Bonanos} {et~al.}(2010){Bonanos}, {Lennon}, {K{\"o}hlinger}, {van
  Loon}, {Massa}, {Sewi{\l}o}, {Evans}, {Panagia}, {Babler}, {Block},
  {Bracker}, {Engelbracht}, {Gordon}, {Hora}, {Indebetouw}, {Meade}, {Meixner},
  {Misselt}, {Robitaille}, {Shiao}, \& {Whitney}}]{ab_smc}
{Bonanos}, A.~Z., {Lennon}, D.~J., {K{\"o}hlinger}, F., {et~al.} 2010, AJ, 140,
  416

\bibitem[{{Bosch} {et~al.}(2009){Bosch}, {Terlevich}, \& {Terlevich}}]{b09}
{Bosch}, G., {Terlevich}, E., \& {Terlevich}, R. 2009, AJ, 137, 3437

\bibitem[{{Bosch} {et~al.}(1999){Bosch}, {Terlevich}, {Melnick}, \&
  {Selman}}]{b99}
{Bosch}, G.~L., {Terlevich}, R., {Melnick}, J., \& {Selman}, F. 1999, A\&AS,
  137, 21

\bibitem[{{Bouwens} {et~al.}(2011){Bouwens}, {Illingworth}, {Labbe}, {Oesch},
  {Trenti}, {Carollo}, {van Dokkum}, {Franx}, {Stiavelli}, {Gonz{\'a}lez},
  {Magee}, \& {Bradley}}]{bouwens11}
{Bouwens}, R.~J., {Illingworth}, G.~D., {Labbe}, I., {et~al.} 2011, Nature,
  469, 504

\bibitem[{{Bresolin} {et~al.}(2006){Bresolin}, {Pietrzy{\'n}ski}, {Urbaneja},
  {Gieren}, {Kudritzki}, \& {Venn}}]{wlm}
{Bresolin}, F., {Pietrzy{\'n}ski}, G., {Urbaneja}, M.~A., {et~al.} 2006, ApJ,
  648, 1007

\bibitem[{{Bresolin} {et~al.}(2007){Bresolin}, {Urbaneja}, {Gieren},
  {Pietrzy{\'n}ski}, \& {Kudritzki}}]{ic1613}
{Bresolin}, F., {Urbaneja}, M.~A., {Gieren}, W., {Pietrzy{\'n}ski}, G., \&
  {Kudritzki}, R. 2007, ApJ, 671, 2028

\bibitem[{{Breysacher}(1981)}]{brey}
{Breysacher}, J. 1981, A\&AS, 43, 203

\bibitem[{{Breysacher} {et~al.}(1999){Breysacher}, {Azzopardi}, \&
  {Testor}}]{bat99}
{Breysacher}, J., {Azzopardi}, M., \& {Testor}, G. 1999, A\&AS, 137, 117

\bibitem[{{Brott} {et~al.}(2011){Brott}, {Evans}, {Hunter}, {de Koter},
  {Langer}, {Dufton}, {Cantiello}, {Trundle}, {Lennon}, {de Mink}, {Yoon}, \&
  {Anders}}]{b11}
{Brott}, I., {Evans}, C.~J., {Hunter}, I., {et~al.} 2011, A\&A, arXiv:1102.0766

\bibitem[{{Brunet} {et~al.}(1975){Brunet}, {Imbert}, {Martin}, {Mianes},
  {Pr{\'e}vot}, {Rebeirot}, \& {Rousseau}}]{bi75}
{Brunet}, J.~P., {Imbert}, M., {Martin}, N., {et~al.} 1975, A\&AS, 21, 109

\bibitem[{{Campbell} {et~al.}(2010){Campbell}, {Evans}, {Mackey}, {Gieles},
  {Alves}, {Ascenso}, {Bastian}, \& {Longmore}}]{mad}
{Campbell}, M.~A., {Evans}, C.~J., {Mackey}, A.~D., {et~al.} 2010, MNRAS, 405,
  421

\bibitem[{{Chen} {et~al.}(2009){Chen}, {Chu}, {Gruendl}, {Gordon}, \&
  {Heitsch}}]{chen09}
{Chen}, C.-H.~R., {Chu}, Y.-H., {Gruendl}, R.~A., {Gordon}, K.~D., \&
  {Heitsch}, F. 2009, ApJ, 695, 511

\bibitem[{{Chu} {et~al.}(1992){Chu}, {Kennicutt}, {Schommer}, \&
  {Laff}}]{cksl92}
{Chu}, Y.-H., {Kennicutt}, R.~C., {Schommer}, R.~A., \& {Laff}, J. 1992, AJ,
  103, 1545

\bibitem[{{Cioni} {et~al.}(2011){Cioni}, {Clementini}, \& {Girardi}}]{vmc}
{Cioni}, M.-R., {Clementini}, G., \& {Girardi}, L. .~{\rm et al}. 2011, A\&A,
  527, A116

\bibitem[{{Crowther} {et~al.}(2010){Crowther}, {Schnurr}, {Hirschi}, {Yusof},
  {Parker}, {Goodwin}, \& {Kassim}}]{c10}
{Crowther}, P.~A., {Schnurr}, O., {Hirschi}, R., {et~al.} 2010, MNRAS, 731

\bibitem[{{Crowther} \& {Smith}(1997)}]{cs97}
{Crowther}, P.~A. \& {Smith}, D.~J. 1997, A\&A, 320, 500

\bibitem[{{de Koter} {et~al.}(1998){de Koter}, {Heap}, \& {Hubeny}}]{dk98}
{de Koter}, A., {Heap}, S.~R., \& {Hubeny}, I. 1998, ApJ, 509, 879

\bibitem[{{de Koter} {et~al.}(2008){de Koter}, {Smith}, \& {Waters}}]{dksw}
{de Koter}, A., {Smith}, L.~J., \& {Waters}, L.~B.~F.~M. 2008, in Mass Loss
  from Stars and the Evolution of Stellar Clusters, ed. {A.~de Koter,
  L.~J.~Smith, \& L.~B.~F.~M.~Waters}, Vol. 388 (Astronomical Society of the
  Pacific)

\bibitem[{{Douglas} {et~al.}(2010){Douglas}, {Bremer}, {Lehnert}, {Stanway}, \&
  {Milvang-Jensen}}]{dbl10}
{Douglas}, L.~S., {Bremer}, M.~N., {Lehnert}, M.~D., {Stanway}, E.~R., \&
  {Milvang-Jensen}, B. 2010, MNRAS, 409, 1155

\bibitem[{{Erb} {et~al.}(2006){Erb}, {Shapley}, {Pettini}, {Steidel}, {Reddy},
  \& {Adelberger}}]{erb06}
{Erb}, D.~K., {Shapley}, A.~E., {Pettini}, M., {et~al.} 2006, ApJ, 644, 813

\bibitem[{{Evans} {et~al.}(2007){Evans}, {Bresolin}, {Urbaneja},
  {Pietrzy{\'n}ski}, {Gieren}, \& {Kudritzki}}]{e07}
{Evans}, C.~J., {Bresolin}, F., {Urbaneja}, M.~A., {et~al.} 2007, ApJ, 659,
  1198

\bibitem[{{Evans} \& {Howarth}(2003)}]{eh03}
{Evans}, C.~J. \& {Howarth}, I.~D. 2003, MNRAS, 345, 1223

\bibitem[{{Evans} {et~al.}(2004){Evans}, {Howarth}, {Irwin}, {Burnley}, \&
  {Harries}}]{eh04}
{Evans}, C.~J., {Howarth}, I.~D., {Irwin}, M.~J., {Burnley}, A.~W., \&
  {Harries}, T.~J. 2004, MNRAS, 353, 601

\bibitem[{{Evans} {et~al.}(2006){Evans}, {Lennon}, {Smartt}, \& {Trundle}}]{f2}
{Evans}, C.~J., {Lennon}, D.~J., {Smartt}, S.~J., \& {Trundle}, C. 2006, A\&A,
  456, 623

\bibitem[{{Evans} {et~al.}(2005){Evans}, {Smartt}, {Lee}, {Lennon}, {Kaufer},
  {Dufton}, {Trundle}, {Herrero}, {Sim{\'o}n-D{\'{\i}}az}, {de Koter},
  {Hamann}, {Hendry}, {Hunter}, {Irwin}, {Korn}, {Kudritzki}, {Langer},
  {Mokiem}, {Najarro}, {Pauldrach}, {Przybilla}, {Puls}, {Ryans}, {Urbaneja},
  {Venn}, \& {Villamariz}}]{e05}
{Evans}, C.~J., {Smartt}, S.~J., {Lee}, J., {et~al.} 2005, A\&A, 437, 467

\bibitem[{{Evans} {et~al.}(2010){Evans}, {Walborn}, {Crowther},
  {H\'enault-Brunet}, {Massa}, {Taylor}, {Howarth}, {Sana}, {Lennon}, \& {van
  Loon}}]{e10}
{Evans}, C.~J., {Walborn}, N.~R., {Crowther}, P.~A., {et~al.} 2010, ApJ, 715,
  L74

\bibitem[{{Feast} {et~al.}(1960){Feast}, {Thackeray}, \& {Wesselink}}]{f60}
{Feast}, M.~W., {Thackeray}, A.~D., \& {Wesselink}, A.~J. 1960, MNRAS, 121, 337

\bibitem[{{Foellmi} {et~al.}(2003){Foellmi}, {Moffat}, \& {Guerrero}}]{f03}
{Foellmi}, C., {Moffat}, A. F.~J., \& {Guerrero}, M.~A. 2003, MNRAS, 338, 1025

\bibitem[{{Gibson}(2000)}]{g00}
{Gibson}, B.~K. 2000, Mem. Soc. Astron. Ital., 71, 693

\bibitem[{{Gieren} {et~al.}(2005){Gieren}, {Pietrzynski}, {Bresolin},
  {Kudritzki}, {Minniti}, {Urbaneja}, {Soszynski}, {Storm}, {Fouque}, {Bono},
  {Walker}, \& {Garcia}}]{araucaria}
{Gieren}, W., {Pietrzynski}, G., {Bresolin}, F., {et~al.} 2005, Msngr, 121, 23

\bibitem[{{Grebel} \& {Chu}(2000)}]{gc00}
{Grebel}, E.~K. \& {Chu}, Y.-H. 2000, AJ, 119, 787

\bibitem[{{Gruendl} \& {Chu}(2009)}]{gc09}
{Gruendl}, R.~A. \& {Chu}, Y.-H. 2009, ApJS, 184, 172 [GC09]

\bibitem[{{Gummersbach} {et~al.}(1995){Gummersbach}, {Zickgraf}, \&
  {Wolf}}]{gzw95}
{Gummersbach}, C.~A., {Zickgraf}, F.-J., \& {Wolf}, B. 1995, A\&A, 302, 409

\bibitem[{{Gvaramadze} \& {Bomans}(2008)}]{gb08}
{Gvaramadze}, V.~V. \& {Bomans}, D.~J. 2008, A\&A, 490, 1071

\bibitem[{{Gvaramadze} {et~al.}(2010){Gvaramadze}, {Kroupa}, \&
  {Pflamm-Altenburg}}]{g10}
{Gvaramadze}, V.~V., {Kroupa}, P., \& {Pflamm-Altenburg}, J. 2010, A\&A, 519,
  A33

\bibitem[{{Gvaramadze} {et~al.}(2011){Gvaramadze}, {Pflamm-Altenburg}, \&
  {Kroupa}}]{g11}
{Gvaramadze}, V.~V., {Pflamm-Altenburg}, J., \& {Kroupa}, P. 2011, A\&A, 525,
  A17

\bibitem[{{Haiman} \& {Loeb}(1997)}]{hl97}
{Haiman}, Z. \& {Loeb}, A. 1997, ApJ, 483, 21

\bibitem[{{Herrero} {et~al.}(2002){Herrero}, {Puls}, \& {Najarro}}]{herrero02}
{Herrero}, A., {Puls}, J., \& {Najarro}, F. 2002, A\&A, 396, 949

\bibitem[{{Herrero} {et~al.}(2000){Herrero}, {Puls}, \&
  {Villamariz}}]{herrero00}
{Herrero}, A., {Puls}, J., \& {Villamariz}, M.~R. 2000, A\&A, 354, 193

\bibitem[{{Hodge}(1988)}]{h88}
{Hodge}, P. 1988, PASP, 100, 1051

\bibitem[{{Howarth} \& {Prinja}(1989)}]{hp89}
{Howarth}, I.~D. \& {Prinja}, R.~K. 1989, ApJS, 69, 527

\bibitem[{{Howarth} {et~al.}(1997){Howarth}, {Siebert}, {Hussain}, \&
  {Prinja}}]{h97}
{Howarth}, I.~D., {Siebert}, K.~W., {Hussain}, G. A.~J., \& {Prinja}, R.~K.
  1997, MNRAS, 284, 265

\bibitem[{{Hunter} {et~al.}(2009){Hunter}, {Brott}, {Langer}, {Lennon},
  {Dufton}, {Howarth}, {Ryans}, {Trundle}, {Evans}, {de Koter}, \&
  {Smartt}}]{ih09}
{Hunter}, I., {Brott}, I., {Langer}, N., {et~al.} 2009, A\&A, 496, 841

\bibitem[{{Hunter} {et~al.}(2008){Hunter}, {Brott}, {Lennon}, {Langer},
  {Dufton}, {Trundle}, {Smartt}, {de Koter}, {Evans}, \& {Ryans}}]{ihapj}
{Hunter}, I., {Brott}, I., {Lennon}, D.~J., {et~al.} 2008, ApJ, 676, L29

\bibitem[{{Hyland} {et~al.}(1992){Hyland}, {Straw}, {Jones}, \& {Gatley}}]{h92}
{Hyland}, A.~R., {Straw}, S., {Jones}, T.~J., \& {Gatley}, I. 1992, MNRAS, 257,
  391

\bibitem[{{Ita} {et~al.}(2010){Ita}, {Onaka}, {Tanab{\'e}}, {Matsunaga},
  {Matsuura}, {Yamamura}, {Nakada}, {Izumiura}, {Ueta}, {Mito}, {Fukushi}, \&
  {Kato}}]{i10}
{Ita}, Y., {Onaka}, T., {Tanab{\'e}}, T., {et~al.} 2010, PASJ, 62, 273

\bibitem[{{Johansson} {et~al.}(1998){Johansson}, {Greve}, {Booth}, {Boulanger},
  {Garay}, {de Graauw}, {Israel}, {Kutner}, {Lequeux}, {Murphy}, {Nyman}, \&
  {Rubio}}]{j98}
{Johansson}, L.~E.~B., {Greve}, A., {Booth}, R.~S., {et~al.} 1998, A\&A, 331,
  857

\bibitem[{{Kato} {et~al.}(2007){Kato}, {Nagashima}, {Nagayama}, {Kurita},
  {Koerwer}, {Kawai}, {Yamamuro}, {Zenno}, {Nishiyama}, {Baba}, {Kadowaki},
  {Haba}, {Hatano}, {Shimizu}, {Nishimura}, {Nagata}, {Sato}, {Murai},
  {Kawazu}, {Nakajima}, {Nakaya}, {Kandori}, {Kusakabe}, {Ishihara},
  {Kaneyasu}, {Hashimoto}, {Tamura}, {Tanab{\'e}}, {Ita}, {Matsunaga},
  {Nakada}, {Sugitani}, {Wakamatsu}, {Glass}, {Feast}, {Menzies}, {Whitelock},
  {Fourie}, {Stoffels}, {Evans}, \& {Hasegawa}}]{irsf}
{Kato}, D., {Nagashima}, C., {Nagayama}, T., {et~al.} 2007, PASJ, 59, 615

\bibitem[{{Kucinskas} {et~al.}(2008){Kucinskas}, {Dobrovolskas},
  {Lazauskait{\.e}}, {Lindegren}, \& {Tanab{\'e}}}]{k08}
{Kucinskas}, A., {Dobrovolskas}, V., {Lazauskait{\.e}}, R., {Lindegren}, L., \&
  {Tanab{\'e}}, T. 2008, BaltA, 17, 283

\bibitem[{{Kudritzki} {et~al.}(2008){Kudritzki}, {Urbaneja}, {Bresolin}, \&
  {Przybilla}}]{kud08}
{Kudritzki}, R.~P., {Urbaneja}, M.~A., {Bresolin}, F., \& {Przybilla}, N. 2008,
  Phys. Scr., T133, 014039

\bibitem[{{Lamers} {et~al.}(1998){Lamers}, {Zickgraf}, {de Winter}, {Houziaux},
  \& {Zorec}}]{l98}
{Lamers}, H. J. G. L.~M., {Zickgraf}, F.-J., {de Winter}, D., {Houziaux}, L.,
  \& {Zorec}, J. 1998, A\&A, 340, 117

\bibitem[{{Landolt}(1992)}]{landolt}
{Landolt}, A.~U. 1992, AJ, 104, 340

\bibitem[{{Langer} {et~al.}(2008){Langer}, {Cantiello}, {Yoon}, {Hunter},
  {Brott}, {Lennon}, {de Mink}, \& {Verheijdt}}]{lcy08}
{Langer}, N., {Cantiello}, M., {Yoon}, S., {et~al.} 2008, in IAU Symp. 250,
  Massive Stars as Cosmic Engines, ed. F.~{Bresolin}, P.~A. {Crowther}, \&
  J.~{Puls} (Cambridge Univ. Press), 167

\bibitem[{{Leitherer} {et~al.}(2010){Leitherer}, {Ortiz Ot{\'a}lvaro},
  {Bresolin}, {Kudritzki}, {Lo Faro}, {Pauldrach}, {Pettini}, \& {Rix}}]{l10}
{Leitherer}, C., {Ortiz Ot{\'a}lvaro}, P.~A., {Bresolin}, F., {et~al.} 2010,
  ApJS, 189, 309

\bibitem[{{Leitherer} {et~al.}(1999){Leitherer}, {Schaerer}, {Goldader},
  {Gonz{\'a}lez Delgado}, {Robert}, {Kune}, {de Mello}, {Devost}, \&
  {Heckman}}]{l99}
{Leitherer}, C., {Schaerer}, D., {Goldader}, J.~D., {et~al.} 1999, ApJS, 123, 3

\bibitem[{{Massey}(2002)}]{m02}
{Massey}, P. 2002, ApJS, 141, 81

\bibitem[{{Massey} \& {Hunter}(1998)}]{mh98}
{Massey}, P. \& {Hunter}, D.~A. 1998, ApJ, 493, 180

\bibitem[{{Massey} {et~al.}(2009){Massey}, {Zangari}, {Morrell}, {Puls},
  {DeGioia-Eastwood}, {Bresolin}, \& {Kudritzki}}]{mfast3}
{Massey}, P., {Zangari}, A.~M., {Morrell}, N.~I., {et~al.} 2009, ApJ, 692, 618

\bibitem[{{Meixner} {et~al.}(2006){Meixner}, {Gordon}, {Indebetouw}, {Hora},
  {Whitney}, {Blum}, {Reach}, {Bernard}, \& {Meade}}]{sage}
{Meixner}, M., {Gordon}, K.~D., {Indebetouw}, R., {et~al.} 2006, AJ, 132, 2268

\bibitem[{{Melnick}(1985)}]{m85}
{Melnick}, J. 1985, A\&A, 153, 235

\bibitem[{{Moffat}(1989)}]{m89}
{Moffat}, A. F.~J. 1989, ApJ, 347, 373

\bibitem[{{Moffat} {et~al.}(1987){Moffat}, {Niemela}, {Philips}, {Chu}, \&
  {Seggewiss}}]{m87}
{Moffat}, A. F.~J., {Niemela}, V.~S., {Philips}, M.~M., {Chu}, Y.-H., \&
  {Seggewiss}, W. 1987, ApJ, 312, 612

\bibitem[{{Mokiem} {et~al.}(2007){Mokiem}, {de Koter}, {Evans}, {Puls},
  {Smartt}, {Crowther}, {Herrero}, {Langer}, {Lennon}, {Najarro}, {Villamariz},
  \& {Vink}}]{rmlmc}
{Mokiem}, M.~R., {de Koter}, A., {Evans}, C.~J., {et~al.} 2007, A\&A, 465, 1003

\bibitem[{{Mokiem} {et~al.}(2006){Mokiem}, {de Koter}, {Evans}, {Puls},
  {Smartt}, {Crowther}, {Herrero}, {Langer}, {Lennon}, {Najarro}, {Villamariz},
  \& {Yoon}}]{rmsmc}
{Mokiem}, M.~R., {de Koter}, A., {Evans}, C.~J., {et~al.} 2006, A\&A, 456, 1131

\bibitem[{{Mokiem} {et~al.}(2005){Mokiem}, {de Koter}, {Puls}, {Herrero},
  {Najarro}, \& {Villamariz}}]{mokiem05}
{Mokiem}, M.~R., {de Koter}, A., {Puls}, J., {et~al.} 2005, A\&A, 441, 711

\bibitem[{{Morgan} \& {Good}(1987)}]{mg87}
{Morgan}, D.~H. \& {Good}, A.~R. 1987, MNRAS, 224, 435

\bibitem[{{Nesvadba} {et~al.}(2008){Nesvadba}, {Lehnert}, {Davies}, {Verma}, \&
  {Eisenhauer}}]{nld08}
{Nesvadba}, N.~P.~H., {Lehnert}, M.~D., {Davies}, R.~I., {Verma}, A., \&
  {Eisenhauer}, F. 2008, A\&A, 479, 67

\bibitem[{{Parker}(1993)}]{p93}
{Parker}, J.~W. 1993, AJ, 106, 560

\bibitem[{{Pasquini} {et~al.}(2002){Pasquini}, {Avila}, {Blecha}, {Cacciari},
  {Cayatte}, {Colless}, {Damiani}, {de Propris}, {Dekker}, {di Marcantonio},
  {Farrell}, {Gillingham}, {Guinouard}, {Hammer}, {Kaufer}, {Hill}, {Marteaud},
  {Modigliani}, {Mulas}, {North}, {Popovic}, {Rossetti}, {Royer}, {Santin},
  {Schmutzer}, {Simond}, {Vola}, {Waller}, \& {Zoccali}}]{flames}
{Pasquini}, L., {Avila}, G., {Blecha}, A., {et~al.} 2002, Msngr, 110, 1

\bibitem[{{Penny}(1996)}]{penny96}
{Penny}, L.~R. 1996, ApJ, 463, 737

\bibitem[{{Penny} \& {Gies}(2009)}]{pg09}
{Penny}, L.~R. \& {Gies}, D.~R. 2009, ApJ, 700, 844

\bibitem[{{Portegies Zwart} {et~al.}(2010){Portegies Zwart}, {McMillan}, \&
  {Gieles}}]{pmg10}
{Portegies Zwart}, S., {McMillan}, S. L.~W., \& {Gieles}, M. 2010, ARA\&A, 48,
  431

\bibitem[{{Prinja} {et~al.}(1990){Prinja}, {Barlow}, \& {Howarth}}]{pbh90}
{Prinja}, R.~K., {Barlow}, M.~J., \& {Howarth}, I.~D. 1990, ApJ, 361, 607

\bibitem[{{Puls} {et~al.}(2011){Puls}, {Sundqvist}, \& {Rivero
  Gonz\'alez}}]{jjj}
{Puls}, J., {Sundqvist}, J.~O., \& {Rivero Gonz\'alez}, J.~G. 2011, in IAU
  Symp. 272, Active OB stars: structure, evolution, mass loss \& critical
  limits, ed. C.~{Neiner}, G.~{Wade}, G.~{Meynet}, \& G.~{Peters} (Cambridge
  Univ. Press), arXiv:1009.0364

\bibitem[{{Puls} {et~al.}(2005){Puls}, {Urbaneja}, {Venero}, {Repolust},
  {Springmann}, {Jokuthy}, \& {Mokiem}}]{p05}
{Puls}, J., {Urbaneja}, M.~A., {Venero}, R., {et~al.} 2005, A\&A, 435, 669

\bibitem[{{Repolust} {et~al.}(2004){Repolust}, {Puls}, \& {Herrero}}]{rp04}
{Repolust}, T., {Puls}, J., \& {Herrero}, A. 2004, A\&A, 415, 349

\bibitem[{{Sana} \& {Evans}(2011)}]{se10}
{Sana}, H. \& {Evans}, C.~J. 2011, in IAU Symp. 272, Active OB stars:
  structure, evolution, mass loss \& critical limits, ed. C.~{Neiner},
  G.~{Wade}, G.~{Meynet}, \& G.~{Peters} (Cambridge Univ. Press),
  arXiv:1009.4197

\bibitem[{{Sanduleak}(1970)}]{sk70}
{Sanduleak}, N. 1970, Contrib. Cerro Tololo Inter-American Obs., No. 89

\bibitem[{{Schild} \& {Testor}(1992)}]{st92}
{Schild}, H. \& {Testor}, G. 1992, A\&AS, 92, 729

\bibitem[{{Schnurr} {et~al.}(2008){Schnurr}, {Moffat}, {St-Louis}, {Morrell},
  \& {Guerrero}}]{schnurr08}
{Schnurr}, O., {Moffat}, A. F.~J., {St-Louis}, N., {Morrell}, N.~I., \&
  {Guerrero}, M.~A. 2008, MNRAS, 389, 806

\bibitem[{{Selman} {et~al.}(1999){Selman}, {Melnick}, {Bosch}, \&
  {Terlevich}}]{s99}
{Selman}, F., {Melnick}, J., {Bosch}, G., \& {Terlevich}, R. 1999, A\&A, 341,
  98

\bibitem[{{Shapley} \& {Lindsay}(1963)}]{sl63}
{Shapley}, H. \& {Lindsay}, E.~M. 1963, IrAJ, 6, 74

\bibitem[{{Skrutskie} {et~al.}(2006){Skrutskie}, {Cutri}, {Stiening},
  {Weinberg}, {Schneider}, {Carpenter}, {Beichman}, {Capps}, {Chester},
  {Elias}, {Huchra}, {Liebert}, {Lonsdale}, {Monet}, {Price}, {Seitzer},
  {Jarrett}, {Kirkpatrick}, {Gizis}, {Howard}, {Evans}, {Fowler}, {Fullmer},
  {Hurt}, {Light}, {Kopan}, {Marsh}, {McCallon}, {Tam}, {Van Dyk}, \&
  {Wheelock}}]{2mass}
{Skrutskie}, M.~F., {Cutri}, R.~M., {Stiening}, R., {et~al.} 2006, AJ, 131,
  1163

\bibitem[{{Smith} {et~al.}(1990){Smith}, {Shara}, \& {Moffat}}]{ssm90}
{Smith}, L.~F., {Shara}, M.~M., \& {Moffat}, A. F.~J. 1990, ApJ, 348, 471

\bibitem[{{Smith} {et~al.}(2010){Smith}, {Povich}, {Whitney}, {Churchwell},
  {Babler}, {Meade}, {Bally}, {Gehrz}, {Robitaille}, \& {Stassun}}]{ns10}
{Smith}, N., {Povich}, M.~S., {Whitney}, B.~A., {et~al.} 2010, MNRAS, 406, 952

\bibitem[{{Stetson}(1987)}]{daophot}
{Stetson}, P.~B. 1987, PASP, 99, 191

\bibitem[{{Stetson}(1994)}]{daomatch}
{Stetson}, P.~B. 1994, PASP, 106, 250

\bibitem[{{Taylor} {et~al.}(2011){Taylor}, {Evans}, \& {Sana}}]{t11}
{Taylor}, W.~D., {Evans}, C.~J., \& {Sana}, H.~{\rm et al}. 2011, A\&A,
  arXiv:1103.5387

\bibitem[{{Testor} {et~al.}(1988){Testor}, {Llebaria}, \& {Debray}}]{tld88}
{Testor}, G., {Llebaria}, A., \& {Debray}, B. 1988, Msngr, 54, 43

\bibitem[{{Testor} \& {Schild}(1990)}]{ts90}
{Testor}, G. \& {Schild}, H. 1990, A\&A, 240, 299

\bibitem[{{Tokunaga} {et~al.}(2002){Tokunaga}, {Simons}, \& {Vacca}}]{tsv02}
{Tokunaga}, A., {Simons}, D.~A., \& {Vacca}, W.~D. 2002, PASP, 114, 180

\bibitem[{{Vaidya} {et~al.}(2009){Vaidya}, {Chu}, {Gruendl}, {Chen}, \&
  {Looney}}]{v09}
{Vaidya}, K., {Chu}, Y.-H., {Gruendl}, R.~A., {Chen}, C.-H.~R., \& {Looney},
  L.~W. 2009, ApJ, 707, 1417

\bibitem[{{Walborn}(1984)}]{w84}
{Walborn}, N.~R. 1984, in IAU Symp. 108, Structure and Evolution of the
  Magellanic Clouds, ed. S.~{van den Bergh} \& K.~S. {de Boer} (Dordrecht:
  Reidel), 243

\bibitem[{{Walborn}(1986)}]{w86}
{Walborn}, N.~R. 1986, in IAU Symp. 116, Luminous Stars and Associations in
  Galaxies, ed. C.~W.~H. {de Loore}, A.~J. {Willis}, \& P.~{Laskarides}
  (Dordrecht: Reidel), 185

\bibitem[{{Walborn}(1991)}]{w91}
{Walborn}, N.~R. 1991, in Massive Stars in Starbursts, ed. C.~{Leitherer},
  N.~{Walborn}, T.~{Heckman}, \& C.~{Norman} (Cambridge University Press), 145

\bibitem[{{Walborn} {et~al.}(1999{\natexlab{a}}){Walborn}, {Barb{\'a}},
  {Brandner}, {Rubio}, {Grebel}, \& {Probst}}]{w99}
{Walborn}, N.~R., {Barb{\'a}}, R.~H., {Brandner}, W., {et~al.}
  1999{\natexlab{a}}, AJ, 117, 225

\bibitem[{{Walborn} \& {Blades}(1997)}]{wb97}
{Walborn}, N.~R. \& {Blades}, J.~C. 1997, ApJS, 112, 457

\bibitem[{{Walborn} {et~al.}(1999{\natexlab{b}}){Walborn}, {Drissen}, {Parker},
  {Saha}, {MacKenty}, \& {White}}]{wd99}
{Walborn}, N.~R., {Drissen}, L., {Parker}, J.~W., {et~al.} 1999{\natexlab{b}},
  AJ, 118, 1684

\bibitem[{{Walborn} \& {Fitzpatrick}(2000)}]{wf00}
{Walborn}, N.~R. \& {Fitzpatrick}, E.~L. 2000, PASP, 112, 50

\bibitem[{{Walborn} {et~al.}(1995){Walborn}, {MacKenty}, {Saha}, {White}, \&
  {Parker}}]{w95}
{Walborn}, N.~R., {MacKenty}, J.~W., {Saha}, A., {White}, R.~L., \& {Parker},
  J.~W. 1995, ApJ, 439, L47

\bibitem[{{Walborn} {et~al.}(2002){Walborn}, {Ma{\'{\i}}z-Apell{\'a}niz}, \&
  {Barb{\'a}}}]{wmb02}
{Walborn}, N.~R., {Ma{\'{\i}}z-Apell{\'a}niz}, J., \& {Barb{\'a}}, R.~H. 2002,
  AJ, 124, 1601

\bibitem[{{Walborn} {et~al.}(2010){Walborn}, {Sota}, {Ma{\'{\i}}z
  Apell{\'a}niz}, {Alfaro}, {Morrell}, {Barb{\'a}}, {Arias}, \& {Gamen}}]{w10}
{Walborn}, N.~R., {Sota}, A., {Ma{\'{\i}}z Apell{\'a}niz}, J., {et~al.} 2010,
  ApJ, 711, L143

\bibitem[{{Werner} {et~al.}(1978){Werner}, {Becklin}, {Gatley}, {Ellis},
  {Hyland}, {Robinson}, \& {Thomas}}]{w78}
{Werner}, M.~W., {Becklin}, E.~E., {Gatley}, I., {et~al.} 1978, MNRAS, 184, 365

\bibitem[{{Whitney} {et~al.}(2008){Whitney}, {Sewilo}, {Indebetouw},
  {Robitaille}, {Meixner}, {Gordon}, {Meade}, {Babler}, {Harris}, {Hora},
  {Bracker}, {Povich}, {Churchwell}, {Engelbracht}, {For}, {Block}, {Misselt},
  {Vijh}, {Leitherer}, {Kawamura}, {Blum}, {Cohen}, {Fukui}, {Mizuno},
  {Mizuno}, {Srinivasan}, {Tielens}, {Volk}, {Bernard}, {Boulanger}, {Frogel},
  {Gallagher}, {Gorjian}, {Kelly}, {Latter}, {Madden}, {Kemper}, {Mould},
  {Nota}, {Oey}, {Olsen}, {Onishi}, {Paladini}, {Panagia}, {Perez-Gonzalez},
  {Reach}, {Shibai}, {Sato}, {Smith}, {Staveley-Smith}, {Ueta}, {Van Dyk},
  {Werner}, {Wolff}, \& {Zaritsky}}]{whitney}
{Whitney}, B.~A., {Sewilo}, M., {Indebetouw}, R., {et~al.} 2008, AJ, 136, 18

\bibitem[{{Zacharias} {et~al.}(2004){Zacharias}, {Urban}, {Zacharias},
  {Wycoff}, {Hall}, {Monet}, \& {Rafferty}}]{ucac2}
{Zacharias}, N., {Urban}, S.~E., {Zacharias}, M.~I., {et~al.} 2004, AJ, 127,
  3043

\bibitem[{{Zaritsky} {et~al.}(2004){Zaritsky}, {Harris}, {Thompson}, \&
  {Grebel}}]{mcps}
{Zaritsky}, D., {Harris}, J., {Thompson}, I.~B., \& {Grebel}, E.~K. 2004, AJ,
  128, 1606

\end{thebibliography}

\onecolumn

\begin{landscape}
{\tiny
\begin{center}
\begin{longtable}{lccccccccll}
  \caption[]{Observational information for the VLT-FLAMES Tarantula
  Survey (VFTS) targets. The FLAMES--Medusa configuration (`A' to `I')
  or FLAMES--ARGUS pointing (`A1' to `A5') used to observe each target
  is listed in the second column; `U' indicates those observed with
  the fibre-feed to UVES.  The fifith and sixth columns give radial
  distances (in arcmin and pc) to the core of R136 (specifically,
  R136-a1: $\alpha$\,$=$\,5$^{\rm h}$\,38$^{\rm
  m}$\,42{\mbox{\ensuremath{.\!\!^{\,\rm s}}}}39,
  $\delta$\,$=$\,$-$69$^{\circ}$\,06$'$\,02\farcs91, J2000.0). Sources
  of photometry are: C (CTIO data, current study); P \citep{p93}; S
  \citep{s99};
    W (WFI data, current study); Z \citep{mcps}.  Aliases/previous identifications of the targets are given in the final column.
    Sources of identifications are: R \citep{f60}; Sk \citep{sk70}; BI
    \citep{bi75}; Brey \citep{brey}; Mk \citep{m85}; M87 \citep{m87};
    T88 \citep{tld88}; ST92 \citep{st92}; P93 \citep{p93}; WB97
    \citep{wb97}; BAT99 \citep{bat99}; S99 \citep{s99}; GC00
    \citep{gc00}; 2MASS \citep{2mass}.  A total of 22 stars have some degree of cross-contamination from
    adjacent spectra on the detector (noted
    in final column), see Section~\ref{crosscont} for details.}\label{cat} \\
\hline
Star & Field & $\alpha$ (2000) & $\delta$ (2000) & \multicolumn{2}{c}{r$_{\rm d}$} & $V$ & $B-V$ & Ref. & Aliases (\& comments) \\
{[VFTS]} & & & & [$'$] & [pc] & & \\
\hline
\endfirsthead
\caption[]{\it{continued}} \\
\hline
Star & Field & $\alpha$ (2000) & $\delta$ (2000) & \multicolumn{2}{c}{r$_{\rm d}$} & $V$ & $B-V$ & Ref. & Aliases (\& comments) \\
{[VFTS]} & & & & [$'$] & [pc] & & & \\
\hline
\endhead
\hline
\multicolumn{10}{r}{\it{continued on next page}}\\
\endfoot
\hline
\endlastfoot
\hline
001 & H     & 05 36 53.39 & $-$69 08 18.29 & \oo9.98 &    145.1 & 16.93 & \pp0.05 & W &   \\
002 & C     & 05 36 54.73 & $-$69 11 38.18 &  11.11 &    161.6 & $-$   & $-$     &$-$& HD\,38029a, Sk$-$69$^\circ$~223, Brey 67, BAT99-085; blend of (at least) two components in WFI image\\
003 & B     & 05 36 55.24 & $-$69 11 37.58 &  11.06 &    160.9 & $-$   & $-$     &$-$& HD\,38029b, Sk$-$69$^\circ$~224\\         
004 & I     & 05 36 58.45 & $-$69 06 47.26 & \oo9.30 &    135.2 & 16.87 & \pp0.16 & W &   \\
005 & H     & 05 37 01.74 & $-$69 08 16.23 & \oo9.25 &    134.5 & 16.25 & \pp0.02 & W &   \\
006 & G     & 05 37 02.50 & $-$69 06 43.75 & \oo8.93 &    129.9 & 16.72 & \pp1.78 & W & 2MASS J05370250$-$6906436  \\
007 & H     & 05 37 02.96 & $-$69 02 04.17 & \oo9.72 &    141.4 & 16.74 & \pp0.03 & W &   \\
008 & G     & 05 37 03.11 & $-$69 08 22.57 & \oo9.15 &    133.2 & 16.92 & \pp0.07 & W &   \\
\end{longtable}
\end{center}
}
\end{landscape}

{\tiny
\begin{center}
\begin{longtable}{lcccccccc}
  \caption[]{Near-IR photometry from the InfraRed Survey Facility (IRSF) Magellanic Clouds catalogue \citep{irsf} 
for targets observed by the VLT-FLAMES Tarantula Survey (VFTS).  IRSF photometry is not available for
six Medusa targets: VFTS~151, 503, 275, 620, 823, and 828; nor for six ARGUS sources: VFTS~1012, 1014, 1015, 
1019, 1024, and 1025.}\label{irsfcat} \\
\hline
Star & &&&&&&&\\
{[VFTS]} & IRSF identification & $J$ & $\sigma J$ & $H$ & $\sigma H$ & $K_{\rm s}$ & $\sigma K_{\rm s}$ & qflag \\
\hline
\endfirsthead
\caption[]{\it{continued}} \\
\hline
Star & &&&&&&&\\
{[VFTS]} & IRSF identification & $J$ & $\sigma J$ & $H$ & $\sigma H$ & $K_{\rm s}$ & $\sigma K_{\rm s}$ & qflag \\
\hline
\endhead
\hline
\multicolumn{9}{r}{\it{continued on next page}}\\
\endfoot
\hline
\endlastfoot
\hline
001 & 05365338$-$6908183 & 16.92 & 0.03 & 16.71 & 0.07 & 16.68 & 0.13 & 122 \\
002 & 05365466$-$6911383 & 13.23 & 0.04 &   $-$   &  $-$  &   $-$   &  $-$  & 500 \\
003 & 05365519$-$6911376 & 11.39 & 0.02 & 11.26 & 0.01 & 11.21 & 0.02 & 115 \\
004 & 05365845$-$6906472 & 16.57 & 0.02 & 16.54 & 0.02 & 16.58 & 0.09 & 111 \\
005 & 05370174$-$6908162 & 16.29 & 0.02 & 16.29 & 0.02 & 16.25 & 0.07 & 111 \\
006 & 05370250$-$6906436 & 13.38 & 0.01 & 12.54 & 0.01 & 12.28 & 0.01 & 111 \\
007 & 05370294$-$6902041 & 16.93 & 0.03 & 16.96 & 0.05 & 16.98 & 0.13 & 114 \\
008 & 05370310$-$6908225 & 16.77 & 0.02 & 16.73 & 0.03 & 16.75 & 0.10 & 111 \\
\end{longtable}
\end{center}
}

\twocolumn
\newpage
\appendix
\section{Observational epochs}
{\scriptsize
\begin{table}[h]
\caption[]{Epochs of the Medusa observations of Fields A to D.  Exposure
    times for LR02 and LR03 observations are all 1815\,s (except the
    LR02-05 exposures for Field~D, with an executed exposure time
    of 2000s), with longer exposures at HR15N of
    2265\,s.}\label{mjd_medusa1}
\begin{center}
\begin{tabular}{llll}
\hline
Field & \lam-setting & Epoch & MJD \\
\hline
A & LR02 & 01 [a$+$b]& 54794.187  \\
& LR02 & 02 [a$+$b]& 54794.231  \\
& LR02 & 03 [a$+$b]& 54794.328  \\
& LR02 & 04 [a$+$b]& 54798.303  \\
&      & 04 [c$+$d]& 54836.146  \\
& LR02 & 05 [a$+$b] & 54836.191  \\
& & 05 [c$+$d] & 54867.058 \\
& LR02 & 06 [a$+$b]& 55108.281  \\
& LR03 & 01 [a$+$b]& 54794.279  \\
& LR03 & 02 [a$+$b]& 54795.182  \\
& LR03 & 03 [a$+$b]& 54795.228  \\
& HR15N & 01 [a$+$b]& 54795.281  \\
& HR15N & 02 [a$+$b]& 54796.207  \\
& & & \\
B & LR02 & 01 [a$+$b]& 54804.105  \\
& LR02 & 02 [a$+$b]& 54804.148  \\
& LR02 & 03 [a$+$b]& 54804.191  \\
& LR02 & 04 [a$+$b]& 54836.239  \\
& LR02 & 05 [a$+$b]& 54836.287  \\
&      & 05 [c$+$d]& 54867.109  \\
& LR02 & 06 [a$+$b]& 55114.320  \\
&      & 06 [c] & 55108.318  \\
& LR03 & 01 [a$+$b]& 54808.143  \\
& LR03 & 02 [a$+$b]& 54808.187  \\
& LR03 & 03 [a$+$b]& 54808.230  \\
& HR15N & 01 [a$+$b]& 54804.247  \\
& HR15N & 02 [a$+$b]& 54804.300  \\
& & & \\
C & LR02 & 01 [a$+$b]& 54774.286 \\
& LR02 & 02 [a$+$b]& 54809.098 \\
& LR02 & 03 [a$+$b]& 54809.141 \\
& LR02 & 04 [a$+$b]& 54837.182 \\
& LR02 & 05 [a$+$b]& 54867.159 \\
& LR02 & 06 [a$+$b]& 55112.264 \\
& LR02 & 07 [a$+$b]& 55459.292 \\
& LR03 & 01 [a$+$b]& 54809.191 \\
& LR03 & 02 [a$+$b]& 54809.235 \\
& LR03 & 03 [a$+$b]& 54809.290 \\
& HR15N & 01 [a$+$b]& 54808.285 \\
& HR15N & 02 [a$+$b]& 54819.262 \\
& & & \\
D & LR02 & 01 [a$+$b]& 54824.242  \\
& LR02 & 02 [a$+$b]& 54824.285  \\
& LR02 & 03 [a$+$b]& 54822.156  \\
& LR02 & 04 [a$+$b]& 54858.171  \\
& LR02 & 05 [a$+$b]& 54879.052  \\
& LR02 & 06 [a$+$b]& 55111.336  \\
& LR03 & 01 [a$+$b]& 54820.252  \\
& LR03 & 02 [a$+$b]& 54821.328  \\
& LR03 & 03 [a$+$b]& 54822.107  \\
& HR15N & 01 [a$+$b]& 54818.189 \\
& HR15N & 02 [a$+$b]& 54819.062 \\
\hline
\end{tabular}
\end{center}
\end{table}

\begin{table}
\caption[]{Epochs of the Medusa observations of Fields E to I.  Exposure
    times for LR02 and LR03 observations are all 1815\,s, with longer
    exposures at HR15N of 2265\,s.}\label{mjd_medusa2}
\begin{center}
\begin{tabular}{llll}
\hline
Field & \lam-setting & Epoch & MJD \\
\hline
E & LR02 & 01 [a$+$b]& 54822.254 \\
& LR02 & 02 [a$+$b]& 54822.298 \\
& LR02 & 03 [a$+$b] & 54822.341 \\
&      & 03 [c$+$d] & 54824.333 \\
& LR02 & 04 [a$+$b]& 54858.220 \\
& LR02 & 05 [a$+$b]& 54889.052 \\
& LR02 & 06 [a$+$b]& 55114.272 \\
\hline
\end{tabular}
\end{center}
\end{table}
}

\newpage
\begin{table}[h]
\begin{center}
\caption{Epochs of the ARGUS fields (A1-A5). Exposure times for Field~A1 were 525\,s, with six 
exposures per OB (except for epoch \#05, in which exposure times were 2$\times$1815\,s per OB). Exposure
times for the four other fields are all 1815\,s.}\label{mjd_arg1}
\begin{tabular}{llll}
\hline
Field & \lam-setting & Epoch & MJD \\
\hline
A1 & LR02 & 01 [a-f]& 54761.237 \\ 
&      & 02 [a-f]& 54761.287 \\ 
&      & 03 [a-f] & 54845.177 \\ 
&      & 03 [g-l] & 54767.283 \\
 &      & 04 [a-f] & 54876.131 \\ 
&      & 05 [a$+$b] & 55178.167 \\ 
&      & 05 [c$+$d] & 55173.310 \\ 
&      & 05 [e$+$f] & 55172.292 \\  
& & & \\
A2 & LR02 & 01 [a$+$b]& 54790.290 \\ 
&      & 02 [a$+$b]& 54791.140 \\
&      & 03 [a$+$b]& 54846.095 \\
&      & 04 [a$+$b]& 54889.102 \\
&      & 05 [a$+$b]& 55173.232 \\
& & & \\
A3 & LR02 & 01 [a$+$b]& 54791.200 \\
&      & 02 [a$+$b]& 54791.247 \\
&      & 03 [a$+$b]& 54846.256 \\
&      & 04 [a$+$b]& 54890.094 \\
 &    & 05 [a$+$b]& 55202.149 \\  
& & & \\
A4 & LR02 & 01 [a$+$b]& 54791.317 \\
&      & 02 [a$+$b]& 54792.176 \\
&      & 03 [a$+$b]& 54847.161 \\
&      & 04 [a$+$b] & 54898.057 \\ 
&      & 04 [c$+$d] & 54892.102 \\ 
&      & 04 [e$+$f] & 54894.039 \\ 
&      & 04 [g$+$h] & 54896.040 \\ 
&      & 04 [i] & 54987.020 \\ 
&      & 05 [a$+$b]& 55201.187 \\
& & & \\
A5 & LR02 & 01 [a$+$b]& 54803.172 \\
&      & 02 [a$+$b]& 54803.216 \\
&      & 03 [a$+$b]& 54851.099 \\
&      & 04 [a$+$b] & 54907.085 \\
&      & 04 [c$+$d] & 54893.045 \\  
&      & 04 [e$+$f$+$g] & 54906.014 \\ 
&      & 05 [a$+$b]& 55204.144 \\
\hline
\end{tabular}
\end{center}
\end{table}

\end{document}